\documentclass{aa}  

\usepackage{graphicx}
\usepackage{txfonts,textcomp}
\usepackage{natbib}
\usepackage{placeins}
\usepackage[bookmarks=false, colorlinks=true, citecolor=blue, linkcolor=blue]{hyperref}
\def\kms{\,km\,s$^{-1}$}
\def\ms{\,m\,s$^{-1}$}

\begin{document} 

\title
{A search for the favored hyperfine transition of a 6.7\,GHz methanol maser line}
   \author{A. Kobak
         \inst{1}
         \protect\href{https://orcid.org/0000-0002-1206-9887}{\protect\includegraphics[scale=0.5]{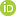}}
          \and
          G. Surcis
          \inst{2} 
          \protect\href{https://orcid.org/0000-0003-4146-9043}{\protect\includegraphics[scale=0.5]{orcid.png}}
          \and
          A. Bartkiewicz
          \inst{1} \protect\href{https://orcid.org/0000-0002-6466-117X}{\protect\includegraphics[scale=0.5]{orcid.png}}
          \and
          W.H.T. Vlemmings
          \inst{3} \protect\href{https://orcid.org/0000-0002-3880-2450}
          {\protect\includegraphics[scale=0.5]{orcid.png}}
          \and
          M. Szymczak
          \inst{1}
          \protect\href{https://orcid.org/0000-0002-1482-8189}
          {\protect\includegraphics[scale=0.5]{orcid.png}}
         }
         
\institute{Institute of Astronomy, Faculty of Physics, Astronomy and Informatics, Nicolaus Copernicus University, Grudziadzka 5, 87-100 Torun, Poland
\and
INAF - Osservatorio Astronomico di Cagliari, Via della Scienza 5, 09047, Selargius, Italy
\and
Department of Space, Earth and Environment, Chalmers University of Technology, 412 96, Gothenburg, Sweden}

  \date{Received  2025 / Accepted 2025}

 
  \abstract
 {The polarized emission of astrophysical masers, especially OH and CH$_3$OH lines, is an effective tool to study the magnetic field in high-mass star-forming regions. The magnetic field strength measurement via the Zeeman effect of OH maser emission is well established, but that of the CH$_3$OH maser emission is still under debate because of its complex hyperfine structure. }
{We aim to identify the dominating hyperfine transition of the Class II 6.7 GHz CH$_3$OH maser emission by comparing the magnetic field strength measured with the 6.0 GHz excited OH maser emission and the Zeeman splitting of the CH$_3$OH maser emission.} 
{We used quasi-simultaneous European VLBI Network observations of the two maser emissions at 6.035\,GHz (excited OH maser) and 6.668\,GHz (CH$_3$OH maser) toward two well-known high-mass young stellar objects: G69.540-0.976 (ON\,1) and G81.871+0.781 (W75N). The observations were performed in full polarimetric mode and in phase-referencing mode to couple the maser features of the two maser emissions in each source.}
{We detected linearly and circularly polarized emission in both maser transitions and high-mass young stellar objects.
Specifically, we measured the magnetic field strength in twelve and five excited OH maser features toward ON\,1 and W75N, respectively, and the Zeeman splitting of the CH$_3$OH maser spectra in one and three maser features toward ON\,1 and W75N, respectively. We determined that the two maser emissions likely probe the same magnetic field but at different densities. Indeed, a direct comparison of the magnetic field strength and the Zeeman splitting as measured with the excited OH and CH$_3$OH maser spots, respectively, provided values of the Zeeman splitting coefficient ($\alpha_{\rm{Z}}$) for the 6.7 GHz CH$_3$OH maser that do not match with any of the table values present in the literature.} 
{We are not able to uniquely identify the dominating hyperfine transition; however, through density considerations we can narrow the choice down to three hyperfine transitions: 3$\rightarrow$4, 6$\rightarrow$7A, and 7$\rightarrow$8. 
Furthermore, we support the previously proposed idea that the favored hyperfine transition is not always the same, but that in different high-mass young stellar objects, the dominating one can be any of these three hyperfine transitions.
}


\titlerunning{Favored hyperfine transition of 6.7 GHz CH$_3$OH maser}
\authorrunning{A. Kobak et al.}

   \maketitle

\section{Introduction}\label{sec:intro}
The magnetic field plays a crucial role in the high-mass star formation process, primarily because it can suppress the fragmentation process and transport angular momentum, keeping the matter close to the central object, where it can be accreted \citep[e.g.,][]{price2007, beuther2018, krumholz2019}.
Observations at high angular resolution of maser emission lines from hydroxyl (OH), water (H$_2$O), or methanol (CH$_3$OH) molecules are the best way to study the kinematics of gas in high-mass young stellar objects \citep[HMYSOs; e.g.,][]{moscadelli2011mas, etoka2012, bartkiewicz2016}, whereas studying their polarized emission allows us to estimate the morphology and strength of magnetic fields (e.g.,~\citealt{bartkiewicz2005, vlemmings2010, surcis2011}).
The pumping mechanism of the maser levels responsible for the CH$_3$OH maser emission at 6.7\,GHz and of the excited OH (hereafter ex-OH) maser emission at 6.035\,GHz is due to the infrared emission from warm dust heated by nearby HMYSOs. In addition, these maser emissions also share similar physical conditions such as temperature and density, making them possibly arise in the same volume of gas \citep{cragg2002}. Therefore, if they arise in the same volume of gas, they must probe the same magnetic field; i.e., the same morphology and strength. 

Hydroxyl (OH) is a paramagnetic molecule; i.e., it strongly interacts with an external magnetic field. This implies that the Zeeman splitting of its maser lines, for typical field strengths in massive star-forming regions, is larger than their linewidths. In other words, the frequency difference between the left- and right-handed circular polarized emissions (LHCP and RHCP), which correspond to the shifted $\sigma$ components produced by the Zeeman effect, is easily measurable. Furthermore, the Landé g-factor and, consequently, the Zeeman splitting coefficient ($\alpha_{\rm{Z}}$) are known for many of the OH maser transitions \citep{davies1974, baudry1997}, making the estimation of the magnetic field strength straightforward. 
Unlike OH, the CH$_3$OH molecule is non-paramagnetic; consequently, the Zeeman splitting of its maser lines is smaller than that of the maser linewidths, making the estimation of the magnetic field strength challenging but feasible \citep[e.g.,][]{vlemmings2010, vlemmings2011}. It is important to note that in this case, the Zeeman splitting is not directly measured, but it must be estimated either by modeling the circularly polarized emission of the masers or by using a cross-correlation method (e.g., \citealt{modjaz2005, vlemmings2008, vlemmings2010}). In addition, all the CH$_3$OH masers show a very complex hyperfine structure \citep{lankhaar2016}. For instance, eight hyperfine transitions can contribute to the maser emission at 6.7\,GHz, and each one has its own Zeeman splitting coefficient \citep{lankhaar2018}. 
At the moment, the dominant hyperfine transition for all of the CH$_3$OH masers is unknown, which prevents the estimation of the magnetic field strength from their Zeeman splitting estimates. Nevertheless, in the case of the 6.7\,GHz CH$_3$OH maser emission the hyperfine transition F=3$\rightarrow$4 has been assumed to be the preferred one \citep[e.g.,][]{lankhaar2018, surcis2022}, although \cite{dallolio2020} suggested that the dominant hyperfine transition might also be either F=6$\rightarrow$7A or F=7$\rightarrow$8. However, the uncertainty on which is the dominating hyperfine transition renders the Zeeman splitting estimates made so far toward a large sample of HMYSOs almost worthless \citep[e.g.,][]{surcis2022}. 
Therefore, it is crucial to find an observational strategy that allows us to identify the dominating hyperfine transition of the 6.7\,GHz CH$_3$OH maser emission. A potential good strategy is that of observing the 6.7\,GHz CH$_3$OH and the 6.035\,GHz ex-OH maser emissions simultaneously toward the same HMYSOs. Indeed, the two maser emissions are thought to arise in the same volume of gas \citep{cragg2002}, and consequently they might probe the same magnetic field. 

In this paper, we present the comparison between the magnetic field strength and the Zeeman splitting as measured and estimated from the circularly polarized emission of the well-known 6.035\,GHz ex-OH and of the 6.7\,GHz CH$_3$OH maser emissions, respectively, with the aim of identifying the dominating hyperfine transition of the latter. The polarized maser emissions have been observed with the European VLBI Network (EVN) near two well-known HMYSOs: G69.540$-$0.976 (ON\,1) and G81.871$+$0.781 (W75N). We selected these two sources according to recent single-dish \citep{szymczak2018, szymczak2020} and interferometric results of a larger sample \citep{kobak2025}.
\cite{kobak2025} observed the two HMYSOs with the
e-MERLIN and showed that in both HMYSOs the CH$_3$OH and ex-OH maser features are very bright and strongly polarized; they shared the same local-standard-of-rest velocity ($V_{\rm{LSR}}$) range, and their projected positions on the plane of the sky coincided. Unfortunately, e-MERLIN results cannot be used for our scientific purposes because the spatial resolution does not ensure the absence of maser line blending, which can severely affect the Zeeman splitting estimates of the  CH$_3$OH maser.
Furthermore, single-dish monitoring observations indicate that both maser emissions are variable on a timescale of one week to a few years \citep[e.g.,][]{szymczak2018, szymczak2020}. Below, we report a brief introduction of the two HMYSOs.

G69.540$-$0.976\footnote{The name follows the Galactic coordinates of the target.} (also known as Onsala\,1, hereafter ON\,1) is a high-mass star-forming region (HMSFR) at a parallactic distance of 2.57$^{+0.34}_{-0.27}$~kpc \citep{rygl2010}. The archival ALMA data (2021.1.00311.S) show that the source hosts three cores of 1.3\,mm dust continuum emission, the westernmost of which is an ultra-compact (UC) H{\sc ii} region that harbors an exciting type B0 star \citep{zheng1985, macleod1998, argon2000, nagayama2008, hu2016} and is associated with 6.7\,GHz CH$_3$OH and 6.035\,GHz ex-OH masers (\citealt{nammahachak2006_on1, sugiyama2011, surcis2022, kobak2025}). 
The other two cores, named WMC1 and WMC2, are associated with 22\,GHz H$_2$O masers \citep{nagayama2008}. 
Two molecular outflows were observed. The first one is traced by H$^{13}$CO$^+$ and SiO and it is oriented on the plane of the sky with a position angle (PA) of $+$44\degr, and the second one is traced by CO \citep[PA=$-$69\degr;][]{kumar2004}. It is still unclear which cores the two outflows are related to.
The 6.7 GHz CH$_3$OH masers show a similar distribution to the 1.665 GHz OH masers, and both maser emissions are tracing an outward motion of the gas, likely suggesting the expansion of the UCH{\sc ii} region \citep{fish2007on1, sugiyama2011}.
Ground-state OH masers (1.665\,GHz and 1.667\,GHz) indicated magnetic field strength along the line of sight ($B_{||}$) ranging from $-$0.9\,mG to $-$5.1\,mG \citep{fish2005}, where a negative $B_{||}$ indicates that the magnetic field is pointing toward the observer. \cite{nammahachak2006_on1} also measured a magnetic field of $-$0.4\,mG and of $-$1\,mG from two Zeeman pairs of 1.720\,GHz OH masers.
Recently, \cite{kobak2025} measured magnetic fields in the $-$1.2\,mG to $-$6.2\,mG range from the 6.035 GHz ex-OH maser, which is consistent with those measured previously with the same ex-OH maser emission ($-1.1$\,mG\,$\leq B_{\rm ||}^{2007}\leq-5.8$\,mG and $-1.2$\,mG\,$\leq B_{\rm ||}^{2010}\leq-12.1$\,mG; \citealt{green2007, fish2010}). 
\cite{green2007} also reported a magnetic field of $-$3.9\,mG from the only Zeeman pair identified toward the OH maser emission at 6.031\,GHz. In addition, they also estimated, using the cross-correlation method, the very first Zeeman splitting ($\Delta V_{\rm{Z}}$) of a 6.7~GHz CH$_3$OH maser --i.e., $\Delta V_{\rm{Z}}=0.9\pm0.3$\,m\,s$^{-1}$-- using the MERLIN.
More recently, \cite{surcis2022} observed the 6.7 GHz CH$_3$OH maser emission with the EVN, and using the full radiative transfer method (FRTM) code they estimated two Zeeman splitting values of 14.5$\pm$4.4\,m\,s$^{-1}$ and 1.2$\pm$0.2\,m\,s$^{-1}$. 

G81.871$+$0.781, known as W75N, is an HMSFR at a parallactic distance of 1.30$\pm$0.07\,kpc \citep{rygl2012}. 
Several radio sources have been identified, among which the most interesting are VLA1 and VLA2 (e.g., \citealt{torrelles1997, carrasco2010, rodriguez2020}). VLA1 is in an early stage of the photoionization, and it is driving a thermal radio jet (PA=$+$42\degr) whose morphology did not change over a period of 18 years \citep{rodriguez2020}. VLA2 is a thermal, collimated, ionized wind surrounded by a dusty disk or envelope, and it varied its morphology between 1996 and 2014 from a compact, roundish source to an elongated source (PA=$+$65\degr; \citealt{carrasco2015}). 
A large-scale CO outflow with PA of $+$66\degr ~was observed by \cite{hunter1994}. Different maser species are associated with VLA1 and VLA2: 6.7\,GHz CH$_3$OH masers (e.g., \citealt{minier2001, surcis2009}), 1.665\,GHz OH masers (e.g., \citealt{hutawarakorn2002, fish2005, fish2011_w75n_oh}), 6.035\,GHz ex-OH masers (e.g., \citealt{kobak2025}), and 22\,GHz H$_2$O masers (e.g., \citealt{surcis2023_w75}). The CH$_3$OH and ex-OH masers are only detected toward VLA1; all the other maser species are detected toward both radio sources.
A magnetic field strength between $+$3.7 and $+$8.1\,mG, indicating that the magnetic field points away from the observer, was measured from the Zeeman effect of a 1.665\,GHz OH maser line around VLA1 \citep{hutawarakorn2002, fish2005},  which is consistent with the values obtained from the Zeeman pair of 6.035\,GHz ex-OH masers ($+2.3$\,mG\,$\leq B_{||}\leq+8.5$\,mG; \citealt{kobak2025}). 
\cite{surcis2009} reported Zeeman splitting estimates, determined via the cross-correlation method, toward three 6.7\,GHz CH$_3$OH maser features, these are $+$0.53\ms, $+$0.75\ms, and $+$0.8\ms. In addition, the magnetic field was measured along VLA1 from the 22\,GHz H$_2$O maser emission at different epochs, indicating typical values in the range $-764$\,mG\,$\leq B_{||}\leq-676$\,mG \citep{surcis2023_w75}. The high values measured from the H$_2$O masers are the consequence of the high densities where these masers arise.

\section{Observations and analysis}
\label{sec:obs}
We observed ON\,1 and W75N with nine antennas of the EVN (Jb, Wb, Ef, Mc, Nt, On, Tr, Ys, Ib) at 6035.092\,MHz and 6668.519\,MHz to detect the ex-OH and CH$_3$OH maser emissions (project code: EK052). To ensure quasi-simultaneous observations of the two maser species, we observed each source on two consecutive days: W75N on 29-30 May, 2023 and ON1 on 31 May-01 June, 2023. The observations were carried out with eight subbands of 4\,MHz ($\sim$100\kms) each, both in phase-referencing (with cycles phase-calibrator - target of 2 min - 3 min) and full-polarization mode, for a total observing time of 48 hours. The phase-referencing calibrators were J200$3+$3034 (at separation 1.7\degr) and J2048$+$4310 (1.9\degr) for ON\,1 and W75N, respectively. The data were correlated with the EVN software correlator (SFXC; \citealt{keimpema2015}) in two correlation passes: a continuum pass for all eight subbands with 128 spectral channels; and a line pass for only the subband and the maser emission with 4096 spectral channels. The line pass allowed us to obtain a spectral resolution of $\sim$1\,kHz (velocity resolution at both frequencies of $\sim$0.05\kms) that is necessary to estimate the Zeeman splitting of the CH$_3$OH maser emission. All four polarization combinations (RR,  LL, RL, and LR) were generated for all correlation passes.

The data were calibrated and imaged by using the Astronomical Image Processing Software package (AIPS, NRAO 2023) following the standard spectral polarimetric and phase-referencing procedures (e.g., \citealt{surcis2012, surcis2023_w75}). For all the datasets, we used the calibrator J2202$+$4216 (BL Lac) to calibrate the bandpass, the delay, the phase, the rates, and the D-terms. The calibration of the polarization angles was instead performed on the calibrator 3C286 ($I\approx0.2$\,Jy\,beam$^{-1}$; linear polarization percentage $P_{\rm{l}}\approx12$\%) for all but one dataset. We could not use 3C286 to calibrate the polarization angles of the CH$_3$OH masers in W75N due to a problem with the 3C286 data. Therefore, we decided to compare the calibrated polarization angle of J2202+4216 from the ON\,1 dataset (PA$=-13$\degr\!$.2\pm2$\degr\!$.1$), which was calibrated using 3C286, with that observed in the W75N dataset, and then applied the rotation to the linear polarization vectors of the CH$_3$OH masers in W75N. The uncertainty of this calibration is equal to 7\degr. We self-calibrated the brightest maser spot of each dataset and applied the solutions to the corresponding dataset. To determine the absolute positions of the maser spots, we performed the phase-referencing calibration between the calibrator and the channel of the brightest maser spot for each dataset. The uncertainties of the absolute positions of the reference brightest maser spot were estimated following \cite{kobak2025}.
Afterward, we imaged the four {\it I}, {\it Q}, {\it U,} and {\it V} Stokes cubes (the latter only for the CH$_3$OH masers) and the {\it RR} and {\it LL} cubes. The {\it Q} and {\it U} cubes were combined to produce cubes of linearly polarized intensity ($POLI=\sqrt{Q^2+U^2}$) and polarization angle ($POLA=0.5 \times \rm{atan}(U/Q)$). The root mean square (rms) noise and the beam size for each imaged cube is reported in Table\,\ref{tab:noise}. 

\begin{table*}
\centering
\scriptsize
\caption{Beam size and the rms noise (1$\sigma$) for each imaged data cube. }
\begin{tabular}{cccccccccc}
\hline
 Source & Restoring & Position & $\sigma_{\mathrm{I}}$ & $\sigma_{\mathrm{Q}}$ & $\sigma_{\mathrm{U}}$ & $\sigma_{\mathrm{V}}$ & $\sigma_{\mathrm{LL}}$ & $\sigma_{\mathrm{RR}}$ & $\sigma_{\mathrm{POLI}}$ \\
  & beam & angle & \\
  & (mas$\times$mas) & (\degr) & (mJy beam$^{-1}$) & (mJy beam$^{-1}$) & (mJy beam$^{-1}$) & (mJy beam$^{-1}$) & (mJy beam$^{-1}$) &  (mJy beam$^{-1}$) & (mJy beam$^{-1}$) \\
\hline
 ON\,1 (CH$_3$OH) & 5.3$\times$4.6 & $-$40 & 4.3 & 5.1 & 5.1 & 4.3 & 7.1 & 4.9 & 3.3\\
 ON\,1 (ex-OH)    & 9.3$\times$5.7 & $-$17 & 1.7 & 9.9 & 9.9 & - & 6.1 & 3.3 & 6.5 \\ 
 W75N (CH$_3$OH)& 5.9$\times$4.3 & $-$47 & 3.7 & 4.1 & 4.4 & 3.8 & 5.5 & 5.2 & 4.5 \\
 W75N (ex-OH)   & 7.3$\times$5.8 & $-$31 & 6.2 & 6.1 & 6.0 & - & 8.5 & 8.3 & 4.1 \\
\hline
\end{tabular}
\label{tab:noise}
\end{table*}

The identification of CH$_3$OH and ex-OH maser features was done as described in \cite{surcis2011}. In particular, the CH$_3$OH maser features were searched in the {\it I} Stokes cube and the ex-OH maser features in the {\it I}, {\it RR}, and {\it LL} cubes. A maser feature is identified when at least three maser spots in consecutive spectral channels are spatially coincident within the beam; therefore, the uncertainty in the identification process is equal to half of the beam. 
We measured the mean linear polarization percentage ($P_{\rm{l}}$) and the mean linear polarization angle ($\chi$) for each identified CH$_3$OH maser feature, considering only the consecutive channels (at least two) across the total intensity spectrum for which the polarized intensity is greater than or equal to $4\sigma_{\rm POLI}$. 

The linearly polarized CH$_3$OH maser features were then analyzed by using the full radiative transfer method (FRTM) code described in \cite{surcis2019}. With this code, we modeled the total intensity ({\it I}) and the {\it POLI} spectra. The outputs of the FRTM code are the emerging brightness temperature ($T_{\rm{b}}\Delta\Omega$, where $\Omega$ is the maser beaming solid angle), the intrinsic maser linewidth ($\Delta V_{\rm{i}}$), and the angle between the magnetic field and the maser propagation direction ($\theta$). If $\theta$ is greater than the Van Vleck angle ($\theta_{\rm{crit}}\approx55\degr$), the magnetic field is perpendicular to the linear polarization angle; otherwise, it is parallel \citep{goldreich1973}.
We note that the FRTM code only works properly for unsaturated masers (e.g., \citealt{vlemmings2010, surcis2011b}). This implies that if the code is performed on a saturated maser, the $\Delta V_{\rm{i}}$ and $T_{\rm{b}}\Delta\Omega$ outputs are overestimated and underestimated, respectively \citep[e.g.,][]{surcis2011b}.
Furthermore, the saturation of the masers introduces a more
complex dependence of the magnetic field with $\theta$ \citep{nedoluha1992}, and consequently the FRTM code provides an overestimate of $\theta$ for these masers, which could also reach values of $90\degr$. A 6.7 GHz CH$_3$OH maser feature can be considered partially saturated if $T_{\rm{b}}\Delta\Omega>2.6 \times10^9$\,K\,sr. However, the orientation of the magnetic field with respect to the linear polarization vectors is not affected by the saturation of the masers \citep{surcis2011b}. Finally, the best estimates of $T_{\rm{b}}\Delta\Omega$ and $\Delta V_{\rm{i}}$ were used to produce {\it I} and {\it V} models with the FRTM code that were used to fit the circularly polarized CH$_3$OH maser feature.
A CH$_3$OH maser feature is considered circularly polarized if the measured {\it V} peak intensity is both >3$\sigma_{\rm V}$ and >3$\sigma_{s.-n.}$, where $\sigma_{s.-n.}$ is the self-noise produced by the maser feature in its channels and becomes important when the power contributed by the astronomical maser is a significant portion of the total received power \citep{sault2012}. From the best {\it I} and {\it V} models, we were able to estimate the circular polarization percentage ($P_{\rm V}$) and $\Delta V_{\rm{Z}}$ of the CH$_3$OH maser features. We note that the FRTM code is able to estimate $\Delta V_{\rm{Z}}$ even if these maser features are partially saturated. These estimates were also made by using the cross-correlation method between the {\it RR} and {\it LL} spectra. 

For each ex-OH maser feature identified in {\it I} cubes, we determined its {\it LL} and {\it RR} counterparts ($>3\sigma_{\rm LL}$ and $>3\sigma_{\rm RR}$) based on their positions on the plane of the sky, to pair them in a so-called Zeeman pair (shifted $\sigma$ components), similarly to what was done in \cite{kobak2025}. The {\it LL} and {\it RR} features were fit with a Gaussian profile that provided the peak intensity and velocity of the maser features. From the velocity difference between the {\it LL} and {\it RR} peaks of a Zeeman pair, we measured $\Delta V_{\rm{Z}}$, and from their and corresponding $I$ intensity peaks we measured $P_{\rm{V}}$. Knowing that the Zeeman coefficient for the ex-OH at 6.035\,GHz is $\alpha_{\rm{Z}}^{\rm{ex-OH}}=0.056\,\rm{ms^{-1}G^{-1}}$ \citep{davies1974,baudry1997}, $B=\Delta V_{\rm{Z}}/\alpha_{\rm{Z}}^{\rm{ex-OH}}$, which is equal to $B_{||}$ if the unshifted $\pi$ component due to the Zeeman effect is negligible. 
We measured $P_{\rm{l}}$ and $\chi$ for each identified ex-OH maser feature, similarly to what we did for the CH$_3$OH maser features, but in this case we considered the channels with polarized intensity greater than or equal to $3\sigma_{\rm POLI}$. For OH masers, and consequently for ex-OH maser emission, the orientation of the magnetic field on the plane of the sky, $\Phi_{\mathrm{B}}$, is perpendicular to the linear polarization vector for $\sigma$ components and parallel to it for the $\pi$ component. As observations have shown in the past, the emission of $\sigma$ components usually dominates over the $\pi$ component, implying that the magnetic field is perpendicular to the linear polarization vector (e.g., \citealt{ gray2003, green2015}). In particular, in about 16\% of the observed cases the $\pi$ component of ex-OH masers is detected, and only in $\sim1\%$ is no intrinsic $\pi$ emission present, suggesting that in all the other cases some unidentified suppression mechanism of the linearly polarized emission ($\pi$ component) must be at play (e.g., \citealt{green2015}). Moreover, the magnetic field is only parallel to the linear polarization vector when $P_{\rm{l}}\geq71\%$. Indeed, only in this case does the unshifted $\pi$ component contribute the most to $P_{\rm{l}}$ (\citealt{fish2006_faraday}). According to our findings (see Sect.~\ref{sec:results}), we can assume that the $\pi$ component ($<3\sigma$) is negligible and, therefore, that the magnetic field is always perpendicular to the linear polarization vectors and we can assume $B\approx B_{||}$ from our Zeeman splitting measurements.

\section{Results}\label{sec:results} 

We detected the 6.7\,GHz CH$_3$OH and 6.035\,GHz ex-OH maser emissions toward both ON\,1 and W75N. The distributions of the identified maser features are shown in Figure\,\ref{fig:on1w75n}, and their parameters are reported in Tables\,\ref{tab:fmethg69}-\ref{tab:fexohw75}. Their total spectra are shown in Figures\,\ref{fig:spectra_on1} and \ref{fig:spectra_w75}. Below, we briefly summarize our results. 

\subsection{G69.540$-$0.976 (ON\,1)} \label{sec:results_on1}

We detected 11 CH$_3$OH (named G69.M01--G69.M11) and 24 ex-OH (named G69.E01--G69.E24) maser features toward ON\,1. Both maser species are spatially distributed in two main groups: one blue-shifted and located in the north of the UCH{\sc ii} region and one red-shifted and located in the south (see Figure\,\ref{fig:on1w75n}). Their distribution and $V_{\rm{LSR}}$ are very similar to those reported previously in \cite{kobak2025}.

Only the brightest CH$_3$OH maser feature (G69.M04, $I$=11\,Jy\,beam$^{-1}$ and $V_{\rm LSR}$=14.68\kms) shows polarized emission; in particular, we measured a linear polarization percentage of $P_{\rm l}$=0.4\% and a circular polarization percentage of $P_{\rm V}$=0.6\%, which is identical to what was measured in 2015 \citep{surcis2022}. Although $P_{\rm l}$ is one third lower than previously measured, the linear polarization angle, $\chi$=$-$58\degr$\pm$6\degr, is similar ($\chi^{\rm 2015}$=$-$34\degr$\pm$9\degr; \citealt{surcis2022}). We were able to properly fit G69.M04 with the FRTM code. The hyperfine transition F=3$\rightarrow$4 was assumed as the preferred one, which provided a value of $\theta$=$+$75\degr$^{+13}_{-37}$, indicating that the orientation of the magnetic field on the plane of the sky is more likely (with a probability of $66\%$ that $\theta>55\degr$) perpendicular to the linear polarization vector (see Sect. \ref{sec:obs}). We note that the FRTM code would provide similar output values, within the errors, for $T_{\rm{b}}\Delta\Omega$, $\Delta V_{\rm{i}}$, and $\theta$  if we assume any other hyperfine transition, this is basically due to the size of our uncertainties. By modeling its {\it V} spectrum (see Figure\,\ref{fig:Vfit}) and by cross-correlating the {\it RR} and {\it LL} spectra, we estimated a Zeeman splitting of 3.2\ms.

We measured linearly polarized emission toward five ex-OH maser features ($3.7\%\leq P_{\rm l}\leq16\%$), with a mean value equal to 7.8\%$\pm$4.95\% and a median value equal to 6.2\%. The linear polarization vectors have position angles ranging from $-$93\degr ~to $-$42\degr ,~with mean and median values of $-$68\degr$\pm$24\degr ~and $-$75\degr, respectively. We measured circularly polarized emission toward all but one (G69.E13) of the ex-OH maser features, but only for 12 of them were we able to identify Zeeman pairs (for them $P_{\rm{V}}\geq77\%$). This implies that the shifted $\sigma$ components dominate over the unshifted $\pi$ component, and consequently the magnetic field is perpendicular to the linear polarization vectors (see Sect.~\ref{sec:obs}). Therefore, $\Phi_{\rm B}$ is oriented northeast-southwest ($\Phi_{\mathrm{B}}^{\rm ex-OH}$=$+$22\degr$\pm$24\degr). Furthermore, we measured the magnetic field along the line of sight for four blue-shifted ($-12.67$\,mG\,$\leq B_{\rm ||}\leq-5.98$\,mG) and eight red-shifted ($-6.37$\,mG\,$\leq B_{\rm ||}\leq-1.35$\,mG) ex-OH maser features.

\subsection{G81.871$+$0.781 (W75N)} \label{sec:results_w75n}

\begin{figure*}
        \includegraphics[width=\textwidth]{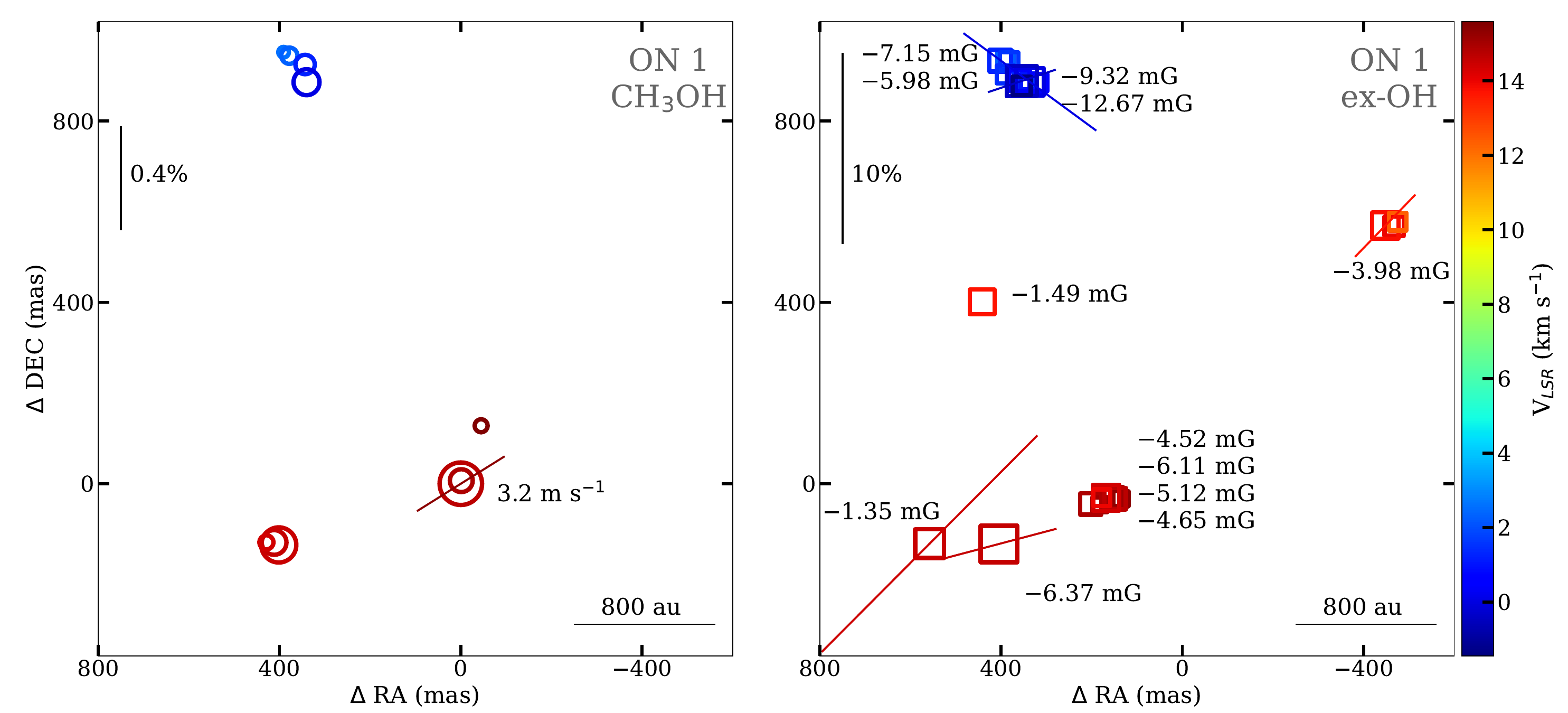}
    \includegraphics[width=\textwidth]{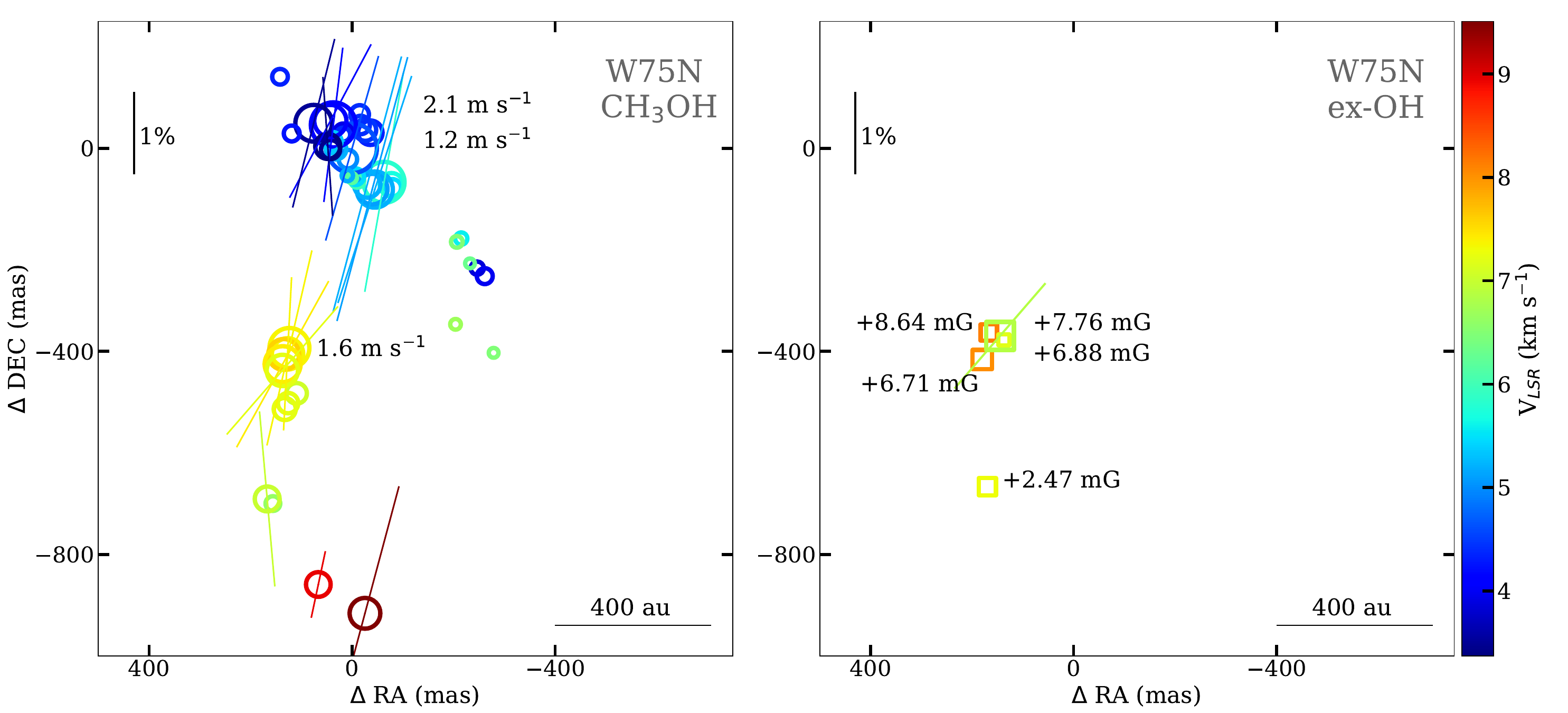}
    \caption{
    Distribution of 6.7\,GHz CH$_3$OH (left panels, circle symbols) and 6.035\,GHz ex-OH (right panel, square symbols) maser features detected around ON\,1 (top panels; $V_{\rm{LSR, sys}}^{\rm{ON\,1}}=+11.6\,\rm{km s^{-1}}$; \citealt{bronfman1996}) and W75N (bottom panels; $V_{\rm{LSR, sys}}^{\rm{W75N}}=+10.0\,\rm{km s^{-1}}$; \citealt{shepherd2003}). The size of the symbols is scaled logarithmically according to their peak intensity (see Tables\,\ref{tab:fmethg69}-\ref{tab:fexohw75}), while their colors indicate the $V_{\rm{LSR}}$ (see Tabes\,\ref{tab:fmethg69}-\ref{tab:fexohw75}). The reference positions are RA(J2000)=20$^{\mathrm{h}}$10$^{\rm{m}}$09\fs04272$\pm$0\fs00007 and Dec(J2000)=$+$31\degr31\arcmin34\farcs9736$\pm$0\farcs0009 for ON\,1; and RA(J2000)=20$^{\mathrm{h}}$38$^{\rm{m}}$36\fs41744$\pm$0\fs00008 and Dec(J2000)=$+$42\degr37\arcmin35\farcs1153$\pm$0\farcs0011 for W75N. The measured linear polarization vectors with their scales, the estimated $\Delta V_{\rm Z}^{CH_3OH}$, and the measured $B_{\rm ||}^{ex-OH}$ are also reported.}
    \label{fig:on1w75n}
\end{figure*}

We identified almost five times the number of 6.7\,GHz CH$_3$OH maser features toward W75N than was previously done by \cite{surcis2009}; i.e., 47 (named here W75.M01-W75N.M47) versus 10. That is due to the high sensitivity of our EVN observations compared to those performed in 2008; indeed, \cite{surcis2009} reported maser features with peak intensities of 2\,Jy\,beam$^{-1}\leq I\leq95.4$\,Jy\,beam$^{-1}$, while we have maser features with 0.03\,Jy\,beam$^{-1}\leq I\leq25.9$\,Jy\,beam$^{-1}$ (see Table\,\ref{tab:fmethw75}). The maser distribution and the $V_{\rm LSR}$ range of the CH$_3$OH maser features are identical to those of 2008; i.e., $+3.37$\,\kms\,$\leq V_{\rm LSR}\leq+9.51$\,\kms ~(see Fig.\,\ref{fig:on1w75n}). We also identified five 6.035\,GHz ex-OH maser features (named W75N.E01-W75N.E05) with $+6.87$\,\kms\,$\leq V_{\rm LSR}\leq+8.13$\,\kms ~that spatially overlap on the plane of the sky with the CH$_3$OH maser features that have a similar velocity range (see Fig.\,\ref{fig:on1w75n} for a direct comparison). All maser features are associated with the radio continuum source VLA\,1.  

\begin{figure*}
\centering
\includegraphics[width = 8 cm]{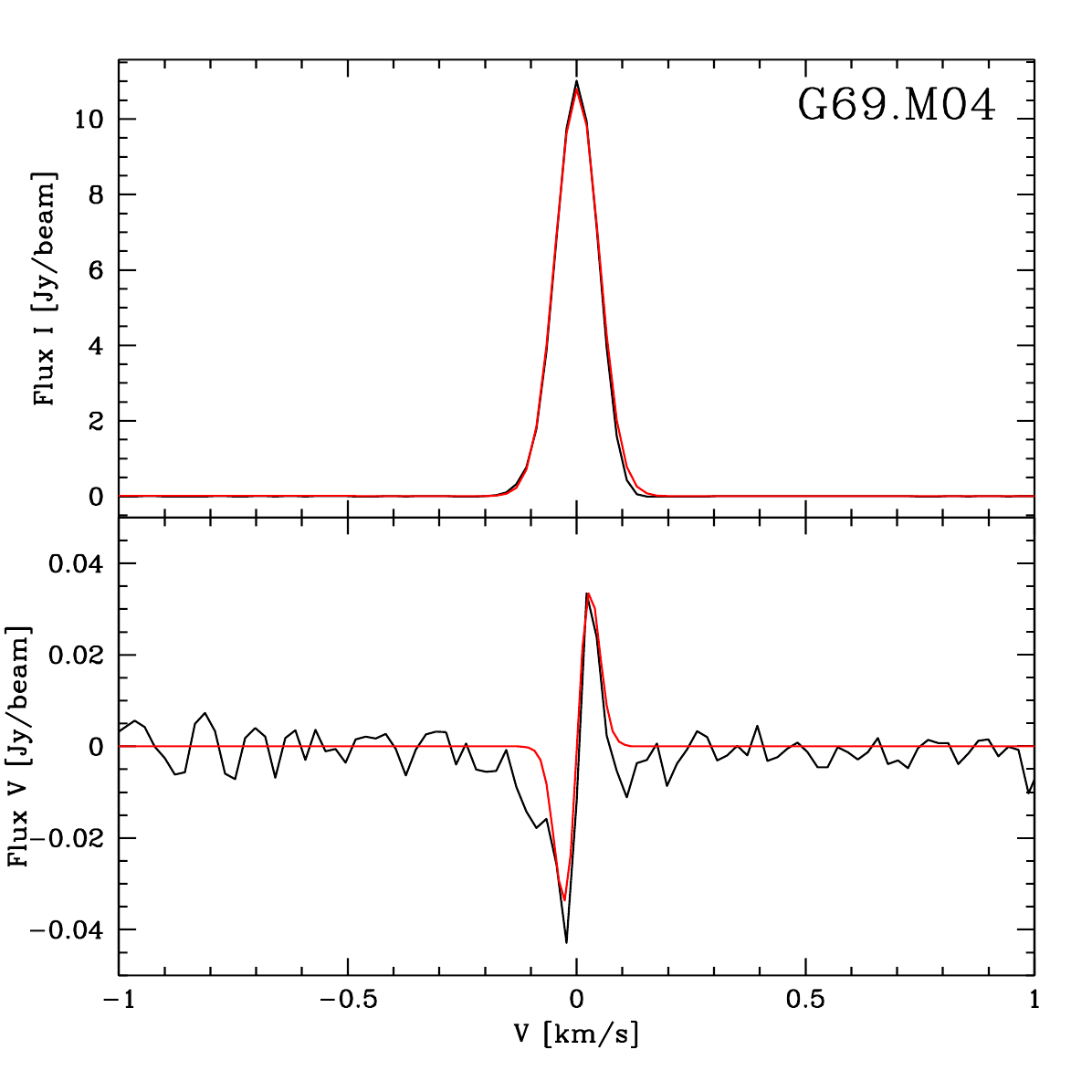}
\includegraphics[width = 8 cm]{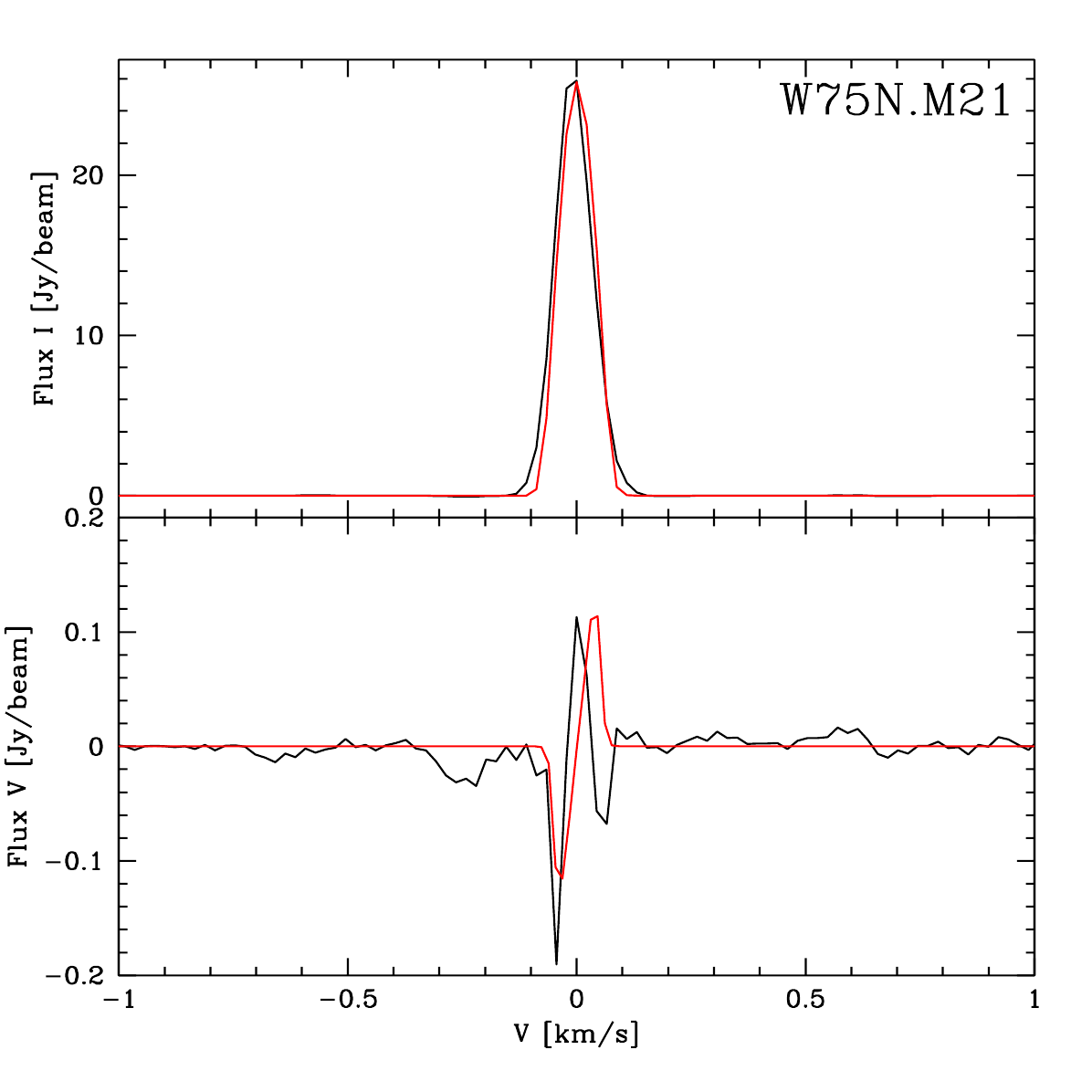}
\includegraphics[width = 8 cm]{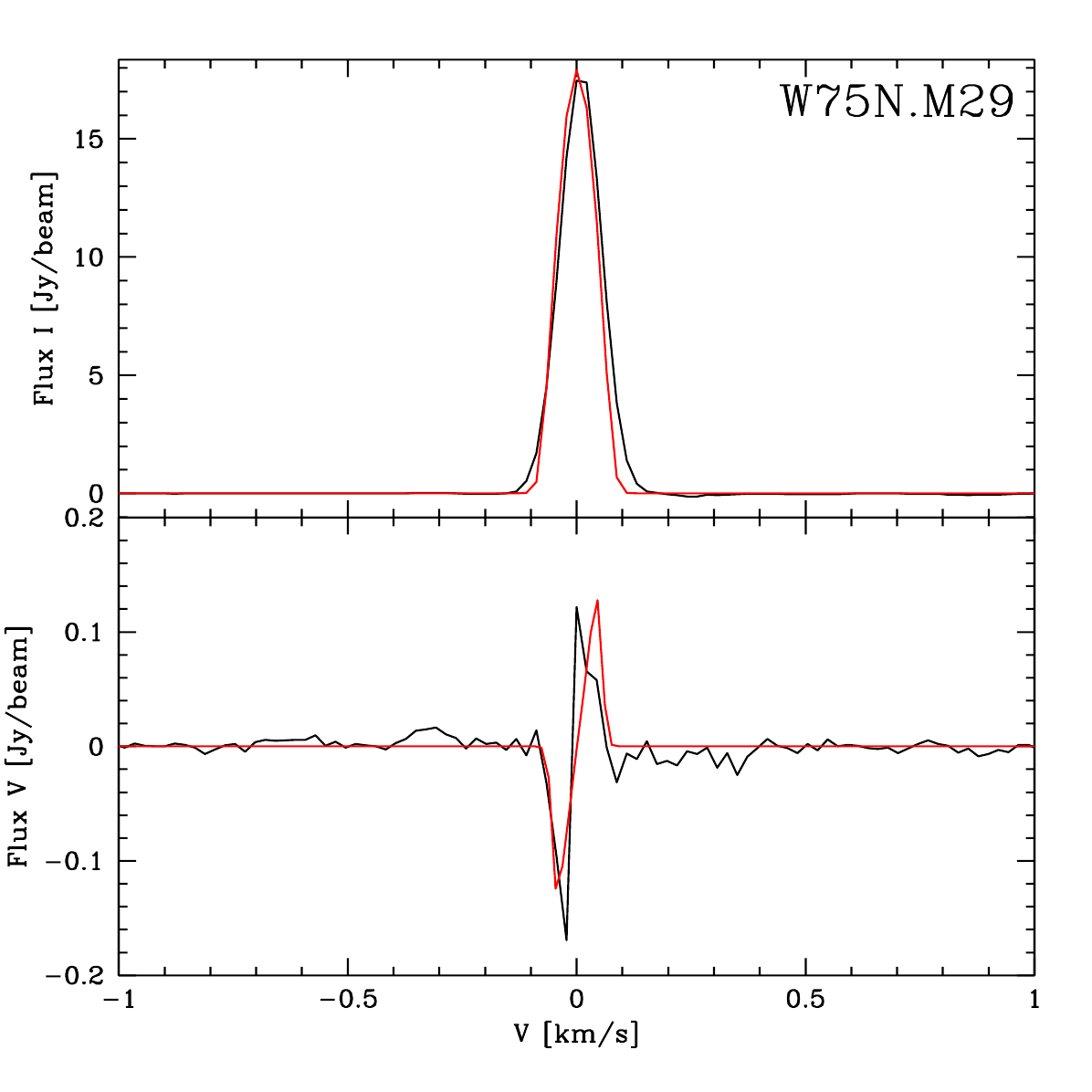}
\includegraphics[width = 8 cm]{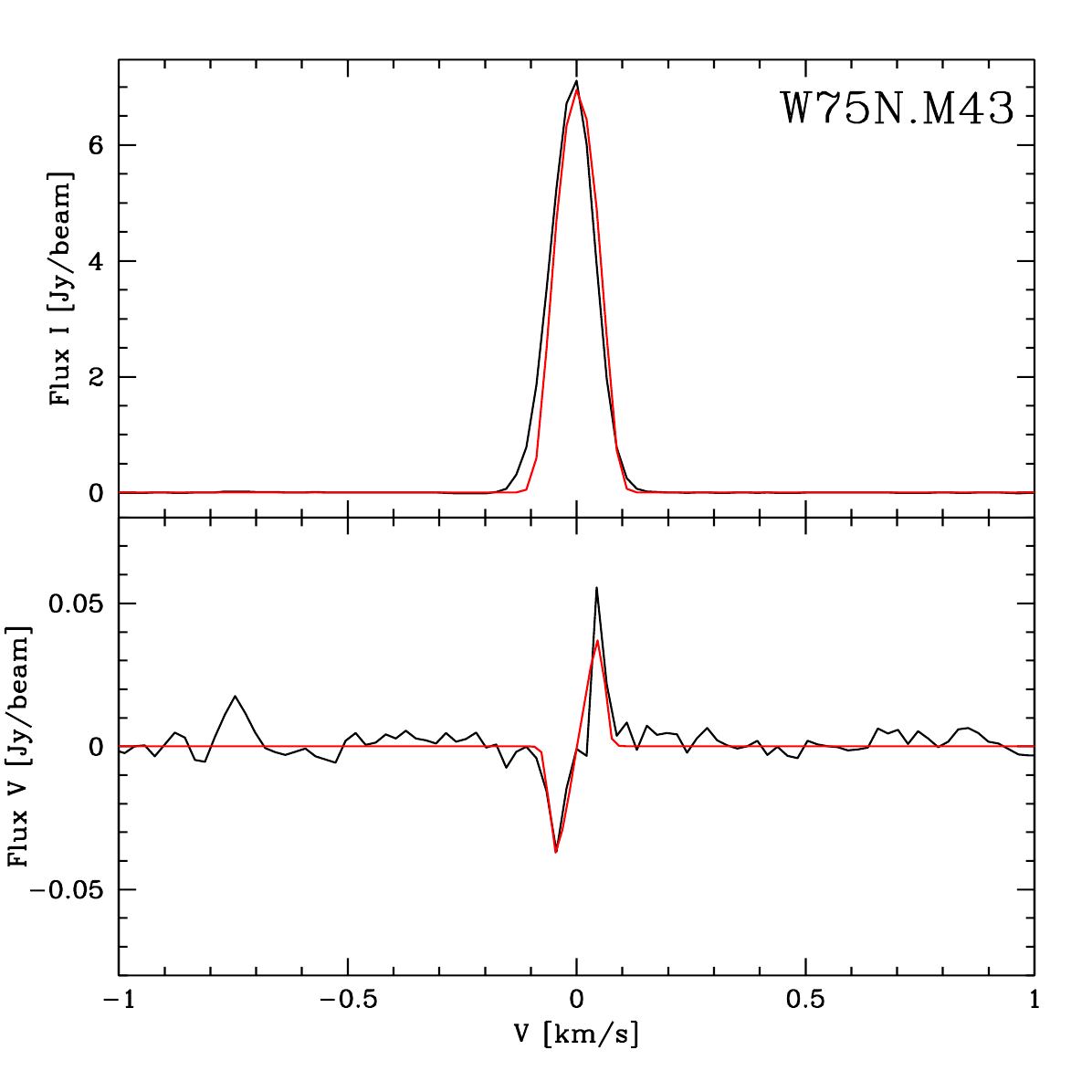}
\caption{Total intensity (\textit{I}, upper panel) and circularly polarized intensity (\textit{V}, lower panel) 
spectra for the 6.7\,GHz $\rm{CH_{3}OH}$ ~maser features named G69.M04, W75N.M21, W75N.M29, and W75N.M43 (see 
Tables~\ref{tab:fmethg69}, and \ref{tab:fmethw75}). The thick red lines are the best-fit models of 
\textit{I} and \textit{V} emissions obtained using the adapted FRTM code (see Sect.\,\ref{sec:obs}). 
The maser features were centered on zero velocity.}
\label{fig:Vfit}
\end{figure*}

We detected linearly polarized emission from 16 CH$_3$OH maser features that show extremely high $P_{\rm l}$, for all but one ($P_{\rm l}^{\rm W75N.M34}$=3.3\%), ranging from 6.8\% to 13.4\%. For comparison, this was $0.9\%\leq P_{\rm l}^{\rm 2008}\leq 4.5\%$ in 2008 \citep{surcis2009}. The linear polarization vectors have position angles between $-$41\degr ~and $+$5\degr, with a mean value of $-$14\degr$\pm$7\degr ~($\chi^{\rm 2008}=-17$\degr$\pm10$\degr, \citealt{surcis2009}). The high percentage of linear polarization suggests either that all the maser features might be saturated and/or that one hyperfine transition is preferred \citep{dallolio2020}. 
However, we report the outputs ($\Delta V_{\rm{i}}$, $T_{\rm{b}}\Delta\Omega$, and $\theta$) of the FRTM code, where we assumed the hyperfine transition F=3$\rightarrow$4 as the preferred one, in Table\,\ref{tab:fmethw75}. The estimated $\theta$ values are all greater than 55\degr ~(see Section\,\ref{sec:obs}); therefore, the magnetic field is perpendicular to the linear polarization vectors of the CH$_3$OH maser features. Actually, all the $\theta$ values are equal to 90\degr,~suggesting that all the maser features might be partially saturated, and therefore the $\theta$ values might be overestimated but still greater than $\theta_{\rm{crit}}\approx55\degr$ (see Sect.~\ref{sec:obs}).
We detected circular polarization toward three CH$_3$OH maser features ($P_{\rm V}^{\rm W75N.M21}=0.9\%$, $P_{\rm V}^{\rm W75N.M29}=1.4\%$, and $P_{\rm V}^{\rm W75N.M43}1.0\%$=). Actually, the detection of $P_{\rm V}^{\rm W75N.M43}$ must be considered as tentative since the {\it V} Stokes emission is about 2$\sigma_{s.-n.}$ (see Fig.\,\ref{fig:Vfit}). From these features, we were able to estimate Zeeman splitting of $\Delta V_{\rm Z}^{\rm W75N.M21}=+1.2$~\ms, $\Delta V_{\rm Z}^{\rm W75N.M29}=+2.1$~\ms, and $\Delta V_{\rm Z}^{\rm W75N.M43}=+1.6$~\ms ~by modeling the {\it V} spectra. The cross-correlation method provided consistent values for W75N.M21 and W75N.M29, but no value was determined for W75N.M43. In 2008, three CH$_3$OH maser features showed $P_{\rm V}\approx0.5\%$ and $+0.5$~\ms~$\leq\Delta V_{\rm Z}\leq+0.8$~\ms \citep{surcis2009}. 

We only measured linearly polarized emission toward the ex-OH maser feature W75N.E02, for which $P_{\rm l}=1.7\%\pm0.6\%$ and $\chi=-36$\degr~$\pm11$\degr, and circularly polarized emission for all the features ($P_{\rm{V}}\geq82\%$). The sky component of the magnetic field is then oriented on the plane of the sky with an angle of $\Phi_{\rm B}^{ex-OH}=+54$\degr~$\pm11$\degr. Furthermore, we identified five Zeeman pairs from which we measured magnetic field strength along the line of sight between $+$2.47\,mG and $+$8.64\,mG. These values are consistent with those obtained with the e-MERLIN observations --i.e., $+2.3$\,mG\,$\leq B_{||}^{\rm{ex-OH, 2025}}\leq +8.5$\,mG-- and reported in \cite{kobak2025}. 

\section{Discussion}
\label{sec:discussion} 
The strategy to identify the most favored hyperfine transition of the 6.7\,GHz CH$_3$OH maser emission is based on the successful calculations of the Zeeman splitting of CH$_3$OH and ex-OH maser lines toward the same volume of gas. This can be done if we assume 
that the magnetic field probed by the two maser species is exactly the same and that its strength is constant within the volume of gas where the masers arise. Therefore, the first step is to determine if the two maser species are probing the same magnetic field (see Sect.\,\ref{sec:orientation}) and then comparing the Zeeman splitting estimates (see Sect.\,\ref{sec:gfactor}). In Sect.\,\ref{sec:comparison_B}, we compare the magnetic fields we measured with those obtained in the past.    

\subsection{magnetic field orientation in ON\,1 and W75N}\label{sec:orientation}
If the CH$_3$OH and ex-OH masers are probing the same magnetic field, we expect that the orientation on the plane of the sky of the magnetic field, as estimated from the linearly polarized emission of the two maser species, is consistent within the same source.
We note that, although we assumed the unshifted $\pi$ ~component of the polarized ex-OH maser features negligible according to our measurements (see Tables~\ref{tab:fexohg69} and \ref{tab:fexohw75}), the magnetic field must not be considered purely parallel to the line of sight. However, only the $\pi$ component can be suppressed by some still-unknown mechanism \citep[e.g.,][]{green2015}. Indeed, from the linear polarization of the $\sigma$ components, we were able to estimate the orientation of the magnetic field on the plane of the sky, which would have been impossible in the case of a magnetic field purely oriented along the line of sight. Therefore, comparing the projected orientation of the magnetic field as estimated from the two maser emissions is still a good approximation.

We measured the magnetic field orientations of $\Phi_{\rm B}^{\rm CH_3OH}=+$32\degr$\pm$6\degr ~and $\langle\Phi_{\rm B}^{\rm ex-OH}\rangle$=$+$22\degr$\pm$24\degr ~in ON\,1. These measurements agree within the uncertainties (see Fig.\,\ref{fig:on1w75n}). The magnetic field in ON\,1 is almost oriented along one of the outflows observed toward the region, particularly the one traced by H$^{13}$CO$^+$ and SiO (PA=+44°). 
Also in the case of W75N the orientation of the magnetic field on the plane of the sky is consistent (see Fig.\,\ref{fig:on1w75n}) between the one estimated from the polarized CH$_3$OH maser emission ($\langle\Phi_{\rm B}^{\rm CH_3OH}\rangle$=$+$76\degr$\pm$7\degr) and that from the ex-OH maser-polarized emission ($\Phi_{\rm B}^{\rm ex-OH}$=$+$54\degr$\pm$11\degr). 
According to our findings, we can assume that the two maser species are likely probing the same magnetic field in each source. 

\subsection{Attempts to identify the favored hyperfine transition for the 6.7\,GHz CH$_3$OH maser emission}\label{sec:gfactor}
Because the 6.7\,GHz CH$_3$OH and 6.035\,GHz ex-OH masers trace a very similar orientation of the magnetic field on the plane of the sky in both sources, we can measure the Zeeman splitting coefficient of the CH$_3$OH maser ($\alpha_{\mathrm{Z}}^{CH_{3}OH}$) using the following relation:

\begin{equation}
    \alpha_\mathrm{Z}^{CH_{3}OH} = \frac{\Delta V_\mathrm{Z}^{CH_{3}OH}}{B_\mathrm{||}^{ex-OH}}\label{eq:alfasplitting}
,\end{equation}

\noindent where $\Delta V_\mathrm{Z}^{\rm CH_3OH}$ is the estimated Zeeman splitting of the CH$_3$OH maser and $B_\mathrm{||}^{ex-OH}$ is the magnetic field strength along the line of sight as measured from the ex-OH maser emission. We note that being able to use $B_\mathrm{||}^{ex-OH}$ rather than $B^{ex-OH}$ reduces the sources of uncertainty. Indeed, from the estimation of $\Delta V_\mathrm{Z}^{\rm CH_3OH}$ it is only possible to measure $B_\mathrm{||}^{CH_3OH}$ \citep[e.g.,][]{vlemmings2008} and not $B^{CH_3OH}=B_\mathrm{||}^{CH_3OH}/\rm{cos} ~\theta$ , for which we would have introduced the large uncertainties in the estimation of $\theta$. Our criteria to couple the ex-OH maser features with the CH$_3$OH maser features are, in order of importance, (1) the measurement of $B_{||}^{ex-OH}$ and estimate of $\Delta V_\mathrm{Z}^{\rm CH_3OH}$; (2) the separation on the plane of the sky, the smaller the better; and (3) a similar $V_{\rm{LSR}}$, the closer the better.

We estimated only one Zeeman splitting toward the 6.7\,GHz CH$_3$OH maser features in ON\,1, i.e., G69.M04 ($V_{\rm LSR}^{\rm G69.M04}$=$+$14.68\kms). We can couple it with G69.E06, which is the closest ex-OH maser feature (with a separation of d=184\,mas, i.e. 470\,au) with similar velocity ($V_{\rm LSR}^{\rm G69.E06}$=$+$15.19\kms), from which we measured magnetic field strength along the line of sight; i.e., $B_{\rm ||}^{\rm G69.E06}$=$-$4.52\,mG. 

In the case of W75N, we estimated the Zeeman splitting toward three CH$_3$OH maser features: W75N.M21, W75N.M29, and W75N.M43 (tentative, see Sect.\,\ref{sec:results_w75n}). Unfortunately, only W75N.M43 ($V_{\rm LSR}^{\rm W75N.M43}$=$+$7.40\kms) is almost coincident with ex-OH maser features (W75N.E01 at d=48\,mas and W75N.E02 at d=182\,mas, i.e., 63\,au and 240\,au, respectively) with similar velocities ($V_{\rm LSR}^{\rm W75N.E01}$=$+$7.20\kms ~and $V_{\rm LSR}^{\rm W75N.E02}$=$+$6.87\kms), for which we measured $B_{||}^{\rm ex-OH}$. 
The other two --i.e., W75N.M21 ($V_{\rm LSR}^{\rm W75N.M21}$=$+$4.64\kms) and W75N.M29 ($V_{\rm LSR}^{\rm W75N.M29}$=$+$4.11\kms)-- are much further from the closest ex-OH maser feature W75N.E02: 367\,mas (480\,au) and 417\,mas (540\,au), respectively.

We list the calculated $\alpha_{\mathrm{Z}}^{\rm CH_3OH}$ values (see Eq.\,\ref{eq:alfasplitting}) in Table\,\ref{tab:alfasplitting} (Col.\,5) for all the pairs of maser features (Col.\,1) mentioned above. The other parameters reported in Table\,\ref{tab:alfasplitting} are the maser features' separation (Col.\,2); the estimated Zeeman splitting of CH$_3$OH's maser emission (Col.\,3); the magnetic field along the line of sight measured from the ex-OH maser features (Col.\,4).
The $\alpha_{\mathrm{Z}}^{\rm CH_3OH}$ obtained from Eq.\,\ref{eq:alfasplitting} do not coincide with any of the Zeeman coefficients of the eight hyperfine transitions of the 6.7\,GHz CH$_3$OH maser emission reported in the supplementary Table\,3 of \cite{lankhaar2018}. This implies that the two maser emissions might not arise in the same volume of gas, or, in other words, that the gas density where they arise is slightly different. Indeed, this condition can still provide the same morphology of the magnetic field, but with a different strength.

\begin{table}
        \centering
    \small
        \caption{Estimated values of the Zeeman splitting coefficient for the 6.7 GHz CH$_3$OH maser emission as calculated from Eq.\,\ref{eq:alfasplitting}.}
        \label{tab:alfasplitting}
        \begin{tabular}{ccccc} 
                \hline
                Maser & $d$ & $\Delta V_{\mathrm{Z}}^{\rm CH_3OH}$ & $B_{\mathrm{||}}^{\rm ex-OH}$ & $\alpha_{\mathrm{Z}}^{\rm CH_3OH}$ \\
          features & (mas) & (m\,s$^{-1}$)  & (mG) & (m\,s$^{-1}$\,G$^{-1}$) \\
                \hline
                G69 (M04, E06) & 184 & 3.2 & $-$4.52 & $-$707.96 \\        
        W75N (M21, E02) & 367 & 2.1 & $+$7.76 & $+$270.62 \\
        W75N (M29, E02) & 417 & 1.2 & $+$7.76 & $+$154.64 \\
        W75N (M43, E02) & 182 & 1.6 & $+$7.76 & $+$206.19\\
        W75N (M43, E01) &  48 & 1.6 & $+$6.88 & $+$232.56 \\        
                \hline
        \end{tabular}
\end{table} 

To compare the number densities of the gas where the CH$_3$OH ($n_{\rm H_2}^{\rm CH_3OH}$) and ex-OH ($n_{\rm H_2}^{\rm ex-OH}$) maser emissions arise, we needed to determine the magnetic field strength ($B_{||}^{\rm CH_3OH}$) from the Zeeman splitting estimates of the 6.7\,GHz CH$_3$OH maser emission. Indeed, following \cite{crutcher2019} we know that
\begin{equation}
\eta \equiv \frac{n_{\rm H_2}^{\rm ex-OH}}{n_{\rm H_2}^{\rm CH_3OH}}=\Biggl(\frac{|B_{||}^{\rm ex-OH}|}{|B_{||}^{\rm CH_3OH}|}\Biggl)^2 \label{eq:rhos}
.\end{equation}
To estimate $B_{||}^{\rm CH_3OH}$, we can assume that one of the eight hyperfine transitions is more favored than the others. As suggested by \cite{lankhaar2018}, this might be F=3$\rightarrow$4 ($\alpha_{\rm Z, 3\rightarrow4}^{\rm CH_3OH}$=$-$50.955\,m\,s$^{-1}$\,G$^{-1}$), because it shows the largest Einstein coefficient. Furthermore, it provides the lowest value of $B_{||}^{\rm CH_3OH}$ \citep{lankhaar2018}. As expected (see Table\,\ref{tab:methB}), the $B_{||}^{\rm CH_3OH}$ values obtained assuming F=3$\rightarrow$4 are larger than those obtained from the associated ex-OH maser features both in ON\,1 and W75N. However, ON\,1 $B_{||}^{\rm CH_3OH}$ and $B_{||}^{\rm ex-OH}$ have the same sign, while they are opposite in W75N for all the maser features. Moreover, while in ON\,1 $\eta$ is about 0.005, in W75N it is at least one order of magnitude larger. Therefore, we can suppose that F=3$\rightarrow$4 might be the favored hyperfine transition in ON\,1, but this is not the case in W75N.
For W75N, we can thus assume that either F=6$\rightarrow$7A ($\alpha_{\rm Z,6\rightarrow7A}^{\rm CH_3OH}$=$+$10.067\,m\,s$^{-1}$\,G$^{-1}$) or F=7$\rightarrow$8 ($\alpha_{\rm Z,7\rightarrow8}^{\rm CH_3OH}$=$+$21.176\,m\,s$^{-1}$\,G$^{-1}$) is the favored one. Indeed, \cite{dallolio2020} suggested these as alternative dominant hyperfine transitions. The results are listed in Table\,\ref{tab:methB}. The sign of $B_{||}^{\rm CH_3OH}$ matches that of $B_{||}^{\rm ex-OH}$ for both hyperfine transitions, and also the $\eta$ values are comparable to what we measure in ON\,1. Nevertheless, it is not possible to identify which of them is the most favorable.
Speculatively, we can state that the two maser emissions arise close to each other, with the 6.7\,GHz CH$_3$OH maser features being closer to the HMYSO, where the gas density is about two orders of magnitude larger than that where the ex-OH maser features arise. This statement can be considered in agreement with the theoretical number-density ranges for the two maser emissions: 10$^5$\,cm$^{-3}<n_{\rm H_2}^{\rm CH_3OH}< 10^9$\,cm$^{-3}$ and 3$\times$10$^6$\,cm$^{-3}<n_{\rm H_2}^{\rm ex-OH}< 3\times10^8$\,cm$^{-3}$ \citep{cragg2002, cragg2005}.

\begin{table*}
        \begin{center}
    \small
        \caption[]{Comparison of the magnetic field along the line of sight as measured from the Zeeman splitting of the 6.7\,GHz CH$_3$OH, considering three different hyperfine transitions, and 6.035\,GHz ex-OH maser features in ON\,1 and W75N.}
        \label{tab:methB}
        \begin{tabular}{ccccccc} 
                \hline
          (1) & (2) & (3) & (4) & (5) & (6) & (7) \\
                Maser & $\Delta V_{\mathrm{Z}}^{\rm CH_3OH}$ & $F_{up} \rightarrow F_{down}$ & $\alpha_{\mathrm{Z}}^{\rm CH_3OH}$ & $B_{\mathrm{||}}^{\rm CH_3OH}$ & $B_{\mathrm{||}}^{\rm ex-OH}$ & $\eta$ \\
          features & & & & & & \\
           & (m\,s$^{-1}$)  & & (m\,s$^{-1}$\,G$^{-1}$) & (mG)  & (mG) & (10$^{-3}$)\\
                \hline
                G69 (M04, E06) & 3.2 &  $3 \rightarrow 4$ & $-$50.955 & $-$62.8 & $-$4.52 & 5.18\\ 
        W75N (M21, E02) & 2.1 &  $3 \rightarrow 4$ & $-$50.955 & $-$41.2 & $+$7.76 & 35.48\\
                         &     &  $6 \rightarrow 7A$ & $+$10.067 & $+$208.6 &  & 1.38\\
                         &     &  $7 \rightarrow 8$ & $+$21.176 & $+$99.2 &  & 6.12\\
        W75N (M29, E02) & 1.2 & $3 \rightarrow 4$ & $-$50.955 & $-$23.6 & $+$7.76 & 108.12\\
                         &     &  $6 \rightarrow 7A$ & $+$10.067 & $+$119.2 &  & 4.24\\
                         &     &  $7 \rightarrow 8$ & $+$21.176 & $+$56.7 & & 18.73\\
        W75N (M43, E02) & 1.6 & $3 \rightarrow 4$ & $-$50.955 & $-$31.4 & $+$7.76 & 61.08\\
                         &     &  $6 \rightarrow 7A$ & $+$10.067 & $+$158.9 & & 2.38\\
                         &     &  $7 \rightarrow 8$ & $+$21.176 & $+$75.6 & & 10.54\\
        W75N (M43, E01) & 1.6 & $3 \rightarrow 4$ & $-$50.955 & $-$31.4 & $+$6.88 & 48.01\\
                         &     &  $6 \rightarrow 7A$ & $+$10.067 & $+$158.9 & & 1.87\\
                         &     &  $7 \rightarrow 8$ & $+$21.176 & $+$75.6 & & 8.29\\
                \hline
        \end{tabular} \end{center}
\end{table*}

\subsection{Comparison of magnetic fields over a long period of time}\label{sec:comparison_B}

Both ON\,1 and W75N were previously searched for polarized maser emissions at 6.035\,GHz (ex-OH) and 6.7\,GHz (CH$_3$OH) to probe their magnetic field morphology and strength. Below, we describe their changes over the years.    

\subsubsection{G69.540-0.976 (ON\,1)}
ON\,1 shows very stable magnetic field morphology over about 20 years. Indeed, \cite{green2007} measured linear polarization vectors for two red-shifted 6.035\,GHz ex-OH maser features ($\chi^{\rm 2007}=-87.7$\degr$\pm2.0$\degr ~and $-42.5$\degr$\pm0.7$\degr), corresponding to our features G69.E22 and G69.E24, and for one red-shifted 6.031\,GHz ex-OH maser feature ($\chi^{\rm 2007}=-43.1$\degr$\pm1.9$\degr), corresponding to our G69.E03, which show the orientation on the plane of the sky to be identical, within the errors, to our measurements. However, we must note that \cite{fish2010} also reported three measurements of linear polarization vectors toward the 6.035\,GHz ex-OH maser features. Two of them --their features K and Z ($\chi^{\rm K,2010}=+52$\degr ~and $\chi^{\rm Z,2010}=-32$\degr)-- can be associated with our features G69.E22 and G69.E03, respectively. Although a direct comparison is difficult because no errors are reported by \cite{fish2010}, taking into account our uncertainties, we have differences of $-$25\degr ~and $-$11\degr, respectively. It is important to emphasize that the linear polarization vectors measured by us toward the ex-OH maser features are all but one (G69.E03, with a difference of $-$12\degr) in perfect agreement with those measured with the e-MERLIN by \cite{kobak2025}. Therefore, for about 20 years the magnetic field, which is always perpendicular to the linear polarization vectors, has been constantly oriented southwest-northeast in the region where the red-shifted maser features arise. Besides the morphology of the magnetic field, we also find that $B_{||}$ did not substantially vary over the same period of time. \cite{green2007}, \cite{fish2010}, and \cite{kobak2025} measured $B_{||}$ toward five, eleven, and five maser features, respectively, with values in the ranges of $-5.8$\,mG~$\leq B_{||}^{\rm{2007}}\leq-1.1$\,mG, $-12.1$\,mG~$\leq B_{||}^{\rm{2010}}\leq-1.3$\,mG, and $-6.2$\,mG~$\leq B_{||}^{\rm{2025}}\leq-1.2$\,mG. By comparing these values with our measurements, we note differences of less than 25\%, which might be due to a high-performance modern instrument, a real magnetic field variation, or both. It is interesting to note that the magnetic field strength measured from the Zeeman effect of ground-state OH maser emissions (1.665\,GHz and 1.667\,GHz) covers a similar range of values: $-3.6$\,mG~$\leq B_{||}^{\rm{OH}}\leq-0.8$\,mG \citep{fish2005,nammahachak2006_on1}.\\
\indent We can compare the magnetic field morphology as probed by the 6.7\,GHz CH$_3$OH masers over eight years. In 2015, \cite{surcis2022} measured linear polarization vectors for two maser features: one blue-shifted located in the north and one red-shifted in the south. The red-shifted one can be associated with our only maser feature that shows linearly polarized emission, i.e., G69.M04. For both maser features the magnetic field is perpendicular to the linear polarization vectors, indicating that the magnetic field is oriented southwest-northeast ($\Phi_{\rm{B}}^{\rm 2022}=+56$\degr$\pm9$\degr ~and $\Phi_{\rm{B}}^{\rm G69.M04}=+32$\degr$\pm6$\degr), which also agrees with the magnetic field orientation probed by the ex-OH maser features in the same red-shifted region. Also, \cite{green2007} estimated linear polarization vectors toward two 6.7\,GHz CH$_3$OH maser features ($\chi^{\rm C, 2007}=+20.6$\degr$\pm2.0$\degr ~and $\chi^{\rm D, 2007}=-76.7$\degr$\pm2.0$\degr), but we cannot perform a direct comparison of the magnetic field morphology because no values of the $\theta$ angles were provided by them. Regardless of the favored hyperfine transition, we can compare our Zeeman splitting estimates with those made in the past. In 2005, \cite{green2007} made one of the first attempts to recover the Zeeman splitting of a 6.7\,GHz CH$_3$OH maser right next to ON\,1. Using the cross-correlation method, they reported a value of $\Delta V_{\mathrm{Z}}^{\rm 2007}=(+0.9\pm0.3$)~$\rm{ms^{-1}}$. Ten years later, \cite{surcis2022} reported, for a feature about 400 mas westward, a value of $\Delta V_{\mathrm{Z}}^{\rm 2022}=(+1.2\pm0.2$)~$\rm{ms^{-1}}$, which was estimated through the FRTM code. This last maser feature can be associated with our maser feature G69.M04, which shows $\Delta V_{\mathrm{Z}}^{\rm G69.M04}=(+3.2\pm0.6$)~$\rm{ms^{-1}}$. From these three estimates made over 18 years and assuming that the hyperfine transition responsible of the observed maser emission was always the same, we can conclude that the magnetic field where the CH$_3$OH maser emission arises might have increased over time. In particular, it might have increased by about 30\% between 2005 and 2015 --if we assume that the magnetic field is uniform in the region-- and 2.6 times between 2015 and 2023, in this case at the exact same location. This behavior is not observed in the magnetic field strength measured from the ex-OH maser emission.

\subsubsection{G81.871+0.781 (W75N)}
Differently from ON\,1, in the case of W75N only the polarized emission of a 6.7\,GHz CH$_3$OH maser was observed in the distant past. Indeed, the previous full polarimetric VLBI observations of the 6.7\,GHz CH$_3$OH and 6.035\,GHz ex-OH maser emissions were performed in 2008 \citep{surcis2009} and 2020 \citep{kobak2025}, respectively. As reported in Sect.~\ref{sec:results_w75n}, the magnetic field morphology and strength as measured from the ex-OH maser features can be considered consistent with what was measured by \cite{kobak2025}, with only a small difference of +6\degr ~for $\Phi_{\rm B}^{\rm ex-OH}$.

The magnetic field, as measured from the linearly polarized CH$_3$OH maser emission, is constantly oriented northeast-southwest. Indeed, \cite{surcis2009} estimated an average magnetic field angle of $\langle\Phi_{\rm B}^{\rm CH_3OH, 2008}\rangle=+73$\degr$\pm$10\degr, ~and we estimated $\langle\Phi_{\rm B}^{\rm CH_3OH}\rangle=+76$\degr$\pm7$\degr. We can even associate each of the eight linearly polarized CH$_3$OH maser features detected by \cite{surcis2009} (A1, A2, A3, A4, A5, B1, C3, and C4) with a maser feature detected by us (W75N.M35, .M29, .M21, .M10, .M12, .M43, .M34, and .M16). Although their peak intensity and $P_{\rm{l}}$ enormously decreased and increased, respectively, in the last 15 years, the local orientation of the magnetic field did not vary. We only notice a negligible difference $<5$\degr ~in three cases: A2, A3, and C3 \citep{surcis2009}. We can only compare the estimated Zeeman splitting for the brightest maser feature detected by \cite{surcis2009}, for which $\Delta V_{\rm Z}^{B1, 2009}=(+0.53\pm0.04)$\ms. Our associated feature --i.e., W75N.M43-- has a value of $\Delta V_{\rm Z}^{\rm W75N.M43}=(+1.6\pm0.3)$\ms, which is three times larger. This suggests an increment of the magnetic field strength in 15 years.

\section{Conclusions}
We observed the polarized emission of 6.7\,GHz CH$_3$OH and 6.035\,GHz ex-OH maser emissions with the EVN toward two HMYSOs: ON\,1 and W75N. The observations at the two frequencies near the same source were performed quasi-simultaneously (only one day apart) to allow the comparison of the magnetic field measured from the ex-OH maser features with the estimated Zeeman splitting of CH$_3$OH maser features;
we also attempted to determine a Zeeman splitting coefficient ($\alpha_{\rm{Z}}^{\rm{CH_3OH}}$) for the CH$_3$OH maser emission to identify the favored hyperfine transition responsible for the 6.7\,GHz CH$_3$OH maser emission. 
We determined that the two maser emissions likely probe the same magnetic field, as we would expect, but at slightly different densities. Indeed, the measured values of $\alpha_{\rm{Z}}^{\rm{CH_3OH}}$ do not match with any of the table values in \cite{lankhaar2018}, implying that we are not able to uniquely identify the dominating hyperfine transition. However, through density considerations, we find that three hyperfine transitions might be responsible for the CH$_3$OH maser emission that we observed toward ON\,1 and W75N. These are 3$\rightarrow$4, 6$\rightarrow$7A, and 7$\rightarrow$8. This might also indicate, as previously suggested \citep[e.g.,][]{dallolio2020}, that the preferred hyperfine transition is not always the same, but that in different HMYSOs this can be any of the three hyperfine transitions we report above. 

Moreover, comparing our magnetic field measurements with those taken decades ago, we found that the magnetic fields toward ON\,1 and W75N did not change their morphology over time. The magnetic field strengths in both sources as probed by the ex-OH maser emission, did not vary over time either. However, we observed a possible increment of the strengths as probed by the CH$_3$OH maser emission, if we assume that the dominating hyperfine transition did not change over time. These differences may be explained by a real variation of the magnetic field closer to the protostar, where the CH$_3$OH masers arise, or by a possible change of the dominating hyperfine transition over time.

In conclusion, we show that the quasi-simultaneous observations of maser emissions from different molecules can be very useful for understanding the physics of the emitting-maser process and of the environment where the masers arise. In our particular case, performing this kind of observation near a large sample of sources might help to statistically identify the most likely hyperfine transition responsible for the 6.7\,GHz CH$_3$OH maser emission. This kind of study will benefit from the future upgrade of existing interferometric networks, such as the ngVLA. In particular, similar studies at higher frequencies could be performed with the EVN thanks to the planned installation of multi-frequency receivers at several stations.

\begin{acknowledgements}
We wish to thank an anonymous referee for making useful suggestions that have improved the paper. The European VLBI Network (www.evlbi.org) is a joint facility of independent European, African, Asian, and North American radio astronomy institutes. Scientific results from data presented in this publication are derived from the following EVN project code EK052.
We acknowledge support from the National Science Centre, Poland, through grant 2021/43/B/ST9/02008.
\end{acknowledgements}

\bibliography{librarian}
\bibliographystyle{aa}
 
\begin{appendix}
  \FloatBarrier
\newpage
\onecolumn
\section{Tables}\label{sec:tables}
\begin{table}[h!]
\caption []{Parameters of the 6.7\,GHz CH$_3$OH maser features detected in G69.540$-$0.976 (ON\,1).} 
\centering
\scriptsize
\begin{tabular}{ l c c c c c c c c c c c c }
\hline
\hline
\,\,\,\,\,(1)&(2)   & (3)      & (4)            & (5)       & (6)              & (7)         & (8)       & (9)                     & (10)                    & (11)                        & (12)         &(13)                           \\
Maser     & RA$^a$ & Dec$^a$ & Peak & $V_{\rm{LSR}}$ & $\Delta v\rm{_{L}}$ & $P_{\rm{l}}^{b}$ &  $\chi^{b}$ & $\Delta V_{\rm{i}}^{c}$ & $T_{\rm{b}}\Delta\Omega^{c}$ & $P_{\rm{V}}$ & $\Delta V_{\rm{Z}}$ & $\theta^{d}$ \\
          &  offset  &  offset  & Intensity ($I$)     &           &                  &             &              &                         &                         &              &                      &      \\
          &  (mas)   &  (mas)   & (Jy beam$^{-1}$)      &  (km s$^{-1}$)   &      (km s$^{-1}$)      & (\%)        &   (\degr)    & (km s$^{-1}$)                  & (log K sr)              &   ($\%$)     &  (m s$^{-1}$)               &    (\degr)  \\
\hline
G69.M01   & -44.763  &  127.548 & $0.100\pm0.005$&  15.65    &      $0.19$      & $-$         & $-$       &  $-$                    & $-$                     & $-$     & $-$                  &$-$ \\
G69.M02   & -9.487   & -4.536   & $0.355\pm0.047$&  14.51    &      $0.19$      & $-$         & $-$       &  $-$                    & $-$                     & $-$     & $-$                  &$-$ \\
G69.M03   & -1.024   &  6.649   & $1.746\pm0.008$&  14.81    &      $0.19$      & $-$         & $-$       &  $-$                    & $-$                     & $-$     & $-$                  &$-$ \\
G69.M04   &  0.000   &  0.000   & $11.022\pm0.049$& 14.68    &      $0.21$      & $0.4\pm0.1$ & $-58\pm6$ &  $0.85^{+0.2}_{-0.3}$   & $8.4^{+1.0}_{-0.5}$     & $0.6$     & $3.2\pm0.6$              &$75^{+13}_{-37}$ \\
G69.M05   &  340.544 &  885.803 & $1.361\pm0.010$&  0.06     &      $0.54$      & $-$         & $-$       &  $-$                    & $-$                     & $-$     & $-$                  &$-$ \\
G69.M06   &  343.715 &  924.751 & $0.438\pm0.006$&  1.25     &      $0.26$      & $-$         & $-$       &  $-$                    & $-$                     & $-$     & $-$                  &$-$ \\
G69.M07   &  378.442 &  944.389 & $0.218\pm0.005$&  2.30     &      $0.25$      & $-$         & $-$       &  $-$                    & $-$                     & $-$     & $-$                  &$-$ \\
G69.M08   &  390.989 &  952.068 & $0.049\pm0.004$&  2.61     &      $0.16$      & $-$         & $-$       &  $-$                    & $-$                     & $-$     & $-$                  &$-$ \\
G69.M09   &  401.805 & -135.044 & $4.730\pm0.047$&  14.51    &      $0.18$      & $-$         & $-$       &  $-$                    & $-$                     & $-$     & $-$                  &$-$ \\
G69.M10   &  411.694 & -129.734 & $0.943\pm0.053$&  14.60    &      $0.20$      & $-$         & $-$       &  $-$                    & $-$                     & $-$     & $-$                  &$-$ \\
G69.M11   &  428.595 & -129.059 & $0.133\pm0.013$&  14.38    &      $0.16$      & $-$         & $-$       &  $-$                    & $-$                     & $-$     & $-$                  &$-$ \\
\hline
\end{tabular} 
\tablefoot{$^a$ The reference position is RA(J2000)=20$^{\mathrm{h}}$10$^{\rm{m}}$09\fs04272$\pm$0\fs00007 and Dec(J2000)=$+$31\degr31\arcmin34\farcs9736$\pm$0\farcs0009. $^b$ $P_{\rm{l}}$ and $\chi$ are the mean values of the linear polarization percentage and the linear polarization angle measured across the spectrum, respectively. $^c$ The best-fitting results obtained by using a model based on the radiative transfer theory of CH$_3$OH masers for $\Gamma+\Gamma_{\nu}=1~\rm{s^{-1}}$ \citep{vlemmings2010, surcis2011}. The errors were determined by analyzing the full probability distribution function. $^d$ The angle between the magnetic field and the maser propagation direction is determined by using the observed $P_{\rm{l}}$ and the fitted emerging brightness temperature. The errors were determined by analyzing the full probability distribution function.} \\
\label{tab:fmethg69}
\end{table}

\begin{table}[h!]
\caption{Parameters of the 6.035\,GHz ex-OH maser features detected in G69.540$-$0.976 (ON\,1).} 
\centering
\scriptsize
\begin{tabular}{ l c c c c c c c c c}
\hline
\hline
\,\,\,\,\,(1)&(2)   & (3)        & (4)           & (5)       & (6)        & (7)  & (8) & (9) & (10)     \\
Maser     & RA$^{a}$ & Dec$^{a}$ & Peak          &  $V_{\rm{LSR}}$  & $P_{\rm{V}}$$^{b}$ & $\Delta V_{\mathrm{Z}}$ & $B_\mathrm{||}$ & $P_{\rm{l}}$ &  $\chi$  \\
          &  offset  &  offset   & Intensity ($I$) &           &        &         \\
          &  (mas)   &  (mas)    & (Jy beam$^{-1}$)     & (km s$^{-1}$) & (\%)   & (km s$^{-1}$) & (mG) & (\%) & ($\degr$)\\
\hline
G69.E01 & $-$475.283 & 577.944 & 0.410 $\pm$ 0.0158 & 12.43 & 97 & $-$0.22 & $-$3.98 & $-$ & $-$\\ 
G69.E02 & $-$467.692 & 568.098 & 0.587 $\pm$ 0.0221 & 13.98 & \textit{23} & $-$ & $-$ & $-$ & $-$\\ 
G69.E03 & $-$448.081 & 569.435 & 2.040 $\pm$ 0.0299 & 13.78 & \textit{100} & $-$ & $-$  & $4.5\pm0.8$ & $-45\pm2$ \\ 
G69.E04 & 135.659 & $-$32.789 & 0.223 $\pm$ 0.0169 & 15.43 & \textit{100} & $-$ & $-$ & $-$ & $-$\\ 
G69.E05 & 148.362 & $-$32.934 & 0.912 $\pm$ 0.0293 & 14.71 & \textit{100} & $-$ & $-$ & $-$ & $-$\\ 
G69.E06 & 154.601 & $-$30.643 & 0.913 $\pm$ 0.021 & 15.19 & 90 & $-$0.25 & $-$4.52 & $-$ & $-$\\ 
G69.E07 & 168.114 & $-$31.119 & 1.948 $\pm$ 0.0462 & 14.37 & 95 & $-$0.34 & $-$6.11 & $-$ & $-$\\ 
G69.E08 & 177.794 & $-$30.877 & 0.340 $\pm$ 0.0199 & 14.03 & 100 & $-$0.29 & $-$5.12 & $-$ & $-$\\ 
G69.E09 & 182.286 & $-$46.535 & 0.223 $\pm$ 0.0175 & 15.09 & \textit{100} &$-$ & $-$ & $-$ & $-$\\ 
G69.E10 & 202.310 & $-$44.848 & 0.875 $\pm$ 0.0197 & 14.85 & 97 & $-$0.26 & $-$4.65 & $-$ & $-$\\ 
G69.E11 & 316.602 & 884.971 & 0.371 $\pm$ 0.0152 & 0.74 & \textit{27} & $-$ & $-$ & $-$ & $-$\\ 
G69.E12 & 336.457 & 886.498 & 2.490 $\pm$ 0.0139 & 0.06 & \textit{35} & $-$ & $-$ & $8.6\pm4.1$ & $+87\pm47$ \\ 
G69.E13 & 339.055 & 887.444 & 0.254 $\pm$ 0.0167 & 1.08 & $-$ & $-$ & $-$ & $-$ & $-$\\ 
G69.E14 & 346.758 & 881.284 & 0.202 $\pm$ 0.0177 & 1.66 & \textit{98} & $-$ & $-$ & $-$ & $-$\\ 
G69.E15 & 354.176 & 888.587 & 3.462 $\pm$ 0.0438 & $-$0.43 & 97 & $-$0.52 & $-$9.31 & $3.7\pm0.9$ & $-75\pm26$ \\ 
G69.E16 & 354.379 & 877.810 & 0.465 $\pm$ 0.0172 & $-$1.45 & 100 & $-$0.71 & $-$12.67 & $-$ & $-$  \\ 
G69.E17 & 359.054 & 889.443 & 0.739 $\pm$ 0.0171 & 0.49 & \textit{84} & $-$ & $-$ & $-$ & $-$\\ 
G69.E18 & 382.130 & 931.563 & 0.471 $\pm$ 0.0176 & 1.61 & \textit{13} & $-$ & $-$ & $-$ & $-$\\ 
G69.E19 & 389.840 & 902.838 & 0.419 $\pm$ 0.017 & 1.46 & 100 & $-$0.40 & $-$7.15 & $-$ & $-$\\ 
G69.E20 & 390.640 & 935.887 & 0.279 $\pm$ 0.0149 & 2.24 & \textit{79} & $-$ & $-$ & $-$ & $-$\\ 
G69.E21 & 401.609 & 933.040 & 0.933 $\pm$ 0.0178 & 1.32 & 77 & $-$0.33 & $-$5.98 & $-$ & $-$\\ 
G69.E22 & 404.653 & $-$132.930 & 7.548 $\pm$ 0.105 & 14.56 & 98 & $-$0.36 & $-$6.37  & $6.2\pm1.4$ & $-85\pm18$ \\ 
G69.E23 & 441.565 & 401.135 & 1.593 $\pm$ 0.0304 & 13.67 & 82 & $-$0.08 & $-$1.49 & $-$ & $-$\\ 
G69.E24 & 557.808 & $-$132.329 & 2.908 $\pm$ 0.0848 & 14.46 & 80 & $-$0.08 & $-$1.35  & $16.0\pm5.1$ & $-42\pm3$\\ 
\hline
\end{tabular}
\tablefoot{$^a$ The reference position is RA(J2000)=20$^{\mathrm{h}}$10$^{\rm{m}}$09\fs04272$\pm$0\fs00007 and Dec(J2000)=$+$31\degr31\arcmin34\farcs9736$\pm$0\farcs0009.
$^b$ The values reported in italics indicate that only one $\sigma$ component has been detected and therefore no $\Delta V_{\mathrm{Z}}$ can be measured.}\\
\label{tab:fexohg69}
\end{table}

\FloatBarrier
\newpage
\begin{table}[h!]
\caption[]{Parameters of the 6.7\,GHz CH$_3$OH maser features detected in G81.871$+$0.781 (W75N).} 
\centering
\scriptsize
\begin{tabular}{ l c c c c c c c c c c c c }
\hline
\hline
\,\,\,\,\,(1)&(2)   & (3)      & (4)            & (5)       & (6)              & (7)         & (8)       & (9)                     & (10)                    & (11)                        & (12)         &(13)                           \\
Maser     & RA$^a$ & Dec$^a$ & Peak  & $V_{\rm{LSR}}$ & $\Delta v\rm{_{L}}$ & $P_{\rm{l}}^{b}$ &  $\chi^{b}$ & $\Delta V_{\rm{i}}^{c}$ & $T_{\rm{b}}\Delta\Omega^{c}$& $P_{\rm{V}}$ & $\Delta V_{\rm{Z}}$  &$\theta^{d}$\\
          &  offset  &  offset  & Intensity ($I$)     &           &                  &             &              &                         &                         &              &                      &      \\
          &  (mas)   &  (mas)   & (Jy beam$^{-1}$)      &  (km s$^{-1}$)   &      (km s$^{-1}$)      & (\%)        &   (\degr)    & (km s$^{-1}$)                  & (log K sr)              &   ($\%$)     &  (m s$^{-1}$)               &(\degr)       \\
\hline
W75N.M01       & -278.842 &  -402.912 & $0.040\pm0.004$& 6.48     &      $0.60$      & $-$         & $-$       &  $-$                    & $-$                     & $-$      & $-$                  &$-$ \\
W75N.M02       & -246.549 &  -236.130 & $0.132\pm0.009$& 3.85     &      $0.38$      & $-$         & $-$       &  $-$                    & $-$                     & $-$      & $-$                  &$-$ \\
W75N.M03       & -232.319 &  -226.574 & $0.039\pm0.004$& 6.31     &      $0.45$      & $-$         & $-$       &  $-$                    & $-$                     & $-$      & $-$                  &$-$ \\
W75N.M04       & -261.454 &  -251.808 & $0.273\pm0.015$& 3.98     &      $0.22$      & $-$         & $-$       &  $-$                    & $-$                     & $-$      & $-$                  &$-$ \\
W75N.M05       & -215.057 &  -177.788 & $0.099\pm0.009$& 5.60     &      $0.26$      & $-$         & $-$       &  $-$                    & $-$                     & $-$      & $-$                  &$-$ \\
W75N.M06       & -206.384 &  -184.334 & $0.083\pm0.004$& 6.48     &      $0.24$      & $-$         & $-$       &  $-$                    & $-$                     & $-$      & $-$                  &$-$ \\
W75N.M07       & -203.816 &  -346.684 & $0.057\pm0.004$& 6.70     &      $0.27$      & $-$         & $-$       &  $-$                    & $-$                     & $-$      & $-$                  &$-$ \\
W75N.M08       & -79.362  &  -77.820  & $0.397\pm0.007$& 5.47     &      $0.22$      & $-$         & $-$       &  $-$                    & $-$                     & $-$      & $-$                  &$-$ \\
W75N.M09       & -77.678  &  -71.812  & $1.299\pm0.015$& 5.69     &      $0.20$      & $-$         & $-$       &  $-$                    & $-$                     & $-$      & $-$                  &$-$ \\
W75N.M10       & -62.985  &  -66.975  & $12.001\pm0.031$& 5.82    &      $0.23$      & $10.9\pm1.5$& $-10\pm1$ &  $0.4^{+0.1}_{-0.2}$    & $10.1^{+0.1}_{-1.0}$    & $-$            & $-$                  &$90^{+7}_{-7}$ \\
W75N.M11       & -44.670  &  -81.238  & $7.285\pm0.028$& 5.21     &      $0.24$      & $11.7\pm0.5$& $-18\pm2$ &  $0.4^{+0.1}_{-0.1}$    & $10.3^{+0.1}_{-0.2}$    & $-$            & $-$                  &$90^{+4}_{-4}$ \\
W75N.M12       & -39.871  &  -80.288  & $3.693\pm0.017$& 5.12     &      $0.25$      & $13.4\pm2.1$& $-15\pm3$ &  $0.5^{+0.2}_{-0.3}$    & $10.2^{+0.2}_{-1.1}$    & $-$            & $-$                  &$90^{+7}_{-7}$ \\
W75N.M13       & -35.997  &   30.689  & $1.530\pm0.014$& 4.33     &      $0.21$      & $-$         & $-$       &  $-$                    & $-$                     & $-$      & $-$                  &$-$ \\
W75N.M14       & -29.808  &  -71.411  & $2.092\pm0.027$& 5.21     &      $0.37$      & $13.0\pm1.2$& $-15\pm8$ &  $0.5^{+0.3}_{-0.1}$    & $10.8^{+0.1}_{-0.2}$    & $-$            & $-$                  &$90^{+8}_{-8}$ \\
W75N.M15       & -25.977  &   36.072  & $0.960\pm0.012$& 4.46     &      $0.19$      & $-$         & $-$       &  $-$                    & $-$                     & $-$      & $-$                  &$-$ \\
W75N.M16       & -25.388  &  -916.046 & $3.827\pm0.011$& 9.51     &      $0.22$      & $12.9\pm0.6$& $-15\pm1$ &  $0.4^{+0.1}_{-0.1}$    & $10.1^{+0.1}_{-0.2}$    & $-$            & $-$                  &$90^{+4}_{-4}$ \\
W75N.M17       & -16.841  &   45.914  & $0.493\pm0.012$& 4.42     &      $0.21$      & $-$         & $-$       &  $-$                    & $-$                     & $-$      & $-$                  &$-$ \\
W75N.M18       & -14.904  &   66.551  & $0.512\pm0.013$& 4.42     &      $0.19$      & $-$         & $-$       &  $-$                    & $-$                     & $-$      & $-$                  &$-$ \\
W75N.M19       & -10.189  &  -63.904  & $0.166\pm0.009$& 6.00     &      $0.23$      & $-$         & $-$       &  $-$                    & $-$                     & $-$      & $-$                  &$-$ \\
W75N.M20       & -7.157   &  -56.038  & $0.322\pm0.008$& 5.47     &      $0.30$      & $-$         & $-$       &  $-$                    & $-$                     & $-$      & $-$                  &$-$ \\
W75N.M21       &  0.000   &   0.000   & $25.899\pm0.051$& 4.64    &      $0.19$      & $9.4\pm0.5$ & $-16\pm1$ &  $0.4^{+0.1}_{-0.1}$    & $10.0^{+0.1}_{-0.1}$    & $0.9$   & $1.2\pm0.2$         &$90^{+5}_{-5}$ \\
W75N.M22       &  0.632   &  -58.525  & $0.047\pm0.004$& 6.18     &      $0.87$      & $-$         & $-$       &  $-$                    & $-$                     & $-$      & $-$                  &$-$ \\
W75N.M23       &  8.294   &  -21.351  & $0.497\pm0.014$& 4.99     &      $1.95$      & $-$         & $-$       &  $-$                    & $-$                     & $-$      & $-$                  &$-$ \\
W75N.M24       &  9.431   &  -53.616  & $0.074\pm0.004$& 5.17     &      $1.98$      & $-$         & $-$       &  $-$                    & $-$                     & $-$      & $-$                  &$-$ \\
W75N.M25       &  17.388  &   30.617  & $0.499\pm0.015$& 3.94     &      $0.89$      & $-$         & $-$       &  $-$                    & $-$                     & $-$      & $-$                  &$-$ \\
W75N.M26       &  17.472  &   27.020  & $0.166\pm0.013$& 4.38     &      $0.23$      & $-$         & $-$       &  $-$                    & $-$                     & $-$      & $-$                  &$-$ \\
W75N.M27       &  31.703  &  -1.568   & $0.760\pm0.028$& 5.21     &      $0.23$      & $-$         & $-$       &  $-$                    & $-$                     & $-$      & $-$                  &$-$ \\
W75N.M28       &  36.797  &   6.535   & $0.334\pm0.015$& 5.30     &      $0.21$      & $-$         & $-$       &  $-$                    & $-$                     & $-$      & $-$                  &$-$ \\
W75N.M29       &  37.008  &   46.204  & $17.471\pm0.047$& 4.11    &      $0.21$      & $7.6\pm0.4$ & $-7\pm4$ &  $0.4^{+0.1}_{-0.1}$    & $10.1^{+0.1}_{-0.2}$    & $1.4$   & $2.1\pm0.3$                 &$90^{+6}_{-6}$ \\
W75N.M30       &  39.534  &   13.592  & $0.570\pm0.016$& 5.34     &      $0.22$      & $-$         & $-$       &  $-$                    & $-$                     & $-$      & $-$                  &$-$ \\
W75N.M31       &  42.481  &  -2.518   & $0.568\pm0.021$& 3.45     &      $0.22$      & $-$         & $-$       &  $-$                    & $-$                     & $-$      & $-$                  &$-$ \\
W75N.M32       &  42.818  &   53.776  & $4.600\pm0.041$& 4.07     &      $0.22$      & $8.5\pm1.2$ & $-28\pm13$&  $0.4^{+0.4}_{-0.1}$    & $10.1^{+0.1}_{-0.6}$    & $-$            & $-$                  &$90^{+9}_{-9}$ \\
W75N.M33       &  47.954  &   3.777   & $1.696\pm0.011$& 3.37     &      $0.19$      & $6.8\pm0.5$ & $+4\pm6$ &  $0.4^{+0.2}_{-0.1}$    & $9.9^{+0.1}_{-0.3}$     & $-$           & $-$                  &$90^{+7}_{-7}$ \\
W75N.M34       &  66.563  &  -859.219 & $1.586\pm0.006$& 8.94     &      $0.19$      & $3.3\pm1.5$ & $-12\pm6$ &  $0.4^{+0.2}_{-0.1}$    & $9.7^{+0.2}_{-0.9}$     & $-$           & $-$                  &$90^{+4}_{-4}$ \\
W75N.M35       &  75.699  &   49.385  & $7.693\pm0.024$& 3.50     &      $0.20$      & $8.5\pm1.5$ & $-14\pm6$ &  $0.4^{+0.1}_{-0.2}$    & $9.9^{+0.1}_{-1.1}$     & $-$           & $-$                  &$90^{+8}_{-8}$ \\
W75N.M36       &  109.339 &  -482.712 & $0.771\pm0.010$& 7.05     &      $0.19$      & $-$         & $-$       &  $-$                    & $-$                     & $-$      & $-$                  &$-$ \\
W75N.M37       &  118.938 &   29.186  & $0.272\pm0.018$& 4.24     &      $0.25$      & $-$         & $-$       &  $-$                    & $-$                     & $-$      & $-$                  &$-$ \\
W75N.M38       &  123.569 &  -393.288 & $10.817\pm0.047$& 7.36    &      $0.20$      & $9.8\pm0.9$ & $-13\pm6$ &  $0.4^{+0.3}_{-0.1}$    & $10.0^{+0.1}_{-0.4}$    & $-$            & $-$                  &$90^{+6}_{-6}$ \\
W75N.M39       &  126.222 &  -501.926 & $0.736\pm0.045$& 7.23     &      $0.19$      & $-$         & $-$       &  $-$                    & $-$                     & $-$      & $-$                  &$-$ \\
W75N.M40       &  127.148 &  -404.903 & $4.089\pm0.048$& 7.36     &      $0.25$      & $7.5\pm1.4$ & $-3\pm9$ &  $0.5^{+0.2}_{-0.3}$    & $10.1^{+0.1}_{-1.0}$    & $-$            & $-$                  &$90^{+11}_{-11}$ \\
W75N.M41       &  132.748 &  -513.611 & $1.014\pm0.025$& 7.23     &      $0.21$      & $-$         & $-$       &  $-$                    & $-$                     & $-$      & $-$                  &$-$ \\
W75N.M42       &  134.811 &  -406.067 & $3.894\pm0.016$& 7.58     &      $0.22$      & $-$         & $-$       &  $-$                    & $-$                     & $-$      & $-$                  &$-$ \\
W75N.M43       &  136.874 &  -425.190 & $7.114\pm0.047$& 7.40     &      $0.22$      & $9.3\pm1.1$ & $-29\pm5$ &  $0.4^{+0.3}_{-0.1}$    & $10.0^{+0.2}_{-0.7}$    & $1.0$          & $1.6\pm0.3$                  &$90^{+7}_{-7}$ \\
W75N.M44       &  137.379 &  -437.584 & $4.103\pm0.025$& 7.23     &      $0.20$      & $8.3\pm3.0$ & $-41\pm11$&  $0.5^{+0.2}_{-0.2}$    & $9.8^{+0.1}_{-1.4}$     & $-$     & $-$                &$90^{+47}_{-47}$ \\
W75N.M45       &  142.094 &   140.930 & $0.258\pm0.014$& 4.33     &      $0.22$      & $-$         & $-$       &  $-$                    & $-$                     & $-$      & $-$                  &$-$ \\
W75N.M46       &  156.030 &  -699.806 & $0.202\pm0.004$& 6.61     &      $2.71$      & $-$         & $-$       &  $-$                    & $-$                     & $-$      & $-$                  &$-$ \\
W75N.M47       &  167.313 &  -690.742 & $1.641\pm0.009$& 7.01     &      $0.22$      & $8.6\pm1.2$ & $+5\pm11$&  $0.4^{+0.3}_{-0.1}$    & $10.1^{+0.1}_{-0.7}$    & $-$            & $-$                  &$90^{+8}_{-8}$ \\
\hline
\end{tabular} 
\tablefoot{$^a$ The reference position is RA(J2000)=20$^{\mathrm{h}}$38$^{\rm{m}}$36\fs41744$\pm$0\fs00008 and Dec(J2000)=$+$42\degr37\arcmin35\farcs1153$\pm$0\farcs0011. $^b$ $P_{\rm{l}}$ and $\chi$ are the mean values of the linear polarization percentage and the linear polarization angle measured across the spectrum, respectively. $^c$ The best-fitting results obtained by using a model based on the radiative transfer theory of CH$_3$OH masers for $\Gamma+\Gamma_{\nu}=1~\rm{s^{-1}}$ \citep{vlemmings2010, surcis2011}. The errors were determined by analyzing the full probability distribution function. $^d$ The angle between the magnetic field and the maser propagation direction is determined by using the observed $P_{\rm{l}}$ and the fitted emerging brightness temperature. The errors were determined by analyzing the full probability distribution function.} \\
\label{tab:fmethw75}
\end{table}

\begin{table}[h!]
\caption{Parameters of the 6.035\,GHz ex-OH maser features detected in G81.871$+$0.781 (W75N).} 
\centering
\scriptsize
\begin{tabular}{ l c c c c c c c c c}
\hline
\hline
\,\,\,\,\,(1)&(2)   & (3)      & (4)            & (5)       & (6)          & (7)  & (8) & (9) & (10)   \\
Maser     & RA$^{a}$ & Dec$^{a}$ & Peak & $V_{\rm{LSR}}$ & $P_{\rm{V}}$ & $\Delta V_{\mathrm{Z}}$ & $B_\mathrm{||}$ & $P_{\rm{l}}$  &  $\chi$\\
          &  offset  &  offset  & Intensity ($I$) &        &        &        &      \\
          &  (mas)   &  (mas)   & (Jy beam$^{-1}$)     & (km s$^{-1}$) & (\%) & (km s$^{-1}$) & (mG) & (\%)& ($\degr$) \\
\hline
W75N.E01 & 137.254 & $-$377.349 & 0.063$\pm$0.007 & 7.20 & 100 & $+$0.39 & $+$6.88 & $-$ & $-$ \\
W75N.E02 & 144.942 & $-$369.462 & 2.596$\pm$0.012 & 6.87 & 84 & $+$0.43 & $+$7.76  & $1.7\pm0.6$ & $-36\pm11$\\
W75N.E03 & 166.882 & $-$362.985 & 0.313$\pm$0.007 & 8.13 & 98 & $+$0.48 & $+$8.64 & $-$ & $-$\\  
W75N.E04 & 169.924 & $-$666.634 & 0.377$\pm$0.007 & 7.30 & 82 & $+$0.14 & $+$2.47 & $-$ & $-$\\
W75N.E05 & 179.807 & $-$416.217 & 0.601$\pm$0.008 & 8.03 & 98 & $+$0.38 & $+$6.71 & $-$ & $-$\\
\hline
\end{tabular}
\tablefoot{$^a$ The reference position is RA(J2000)=20$^{\mathrm{h}}$38$^{\rm{m}}$36\fs41744$\pm$0\fs00009 and Dec(J2000)=$+$42\degr37\arcmin35\farcs1153$\pm$0\farcs0013.}\\ 
\label{tab:fexohw75}
\end{table}

\twocolumn
\FloatBarrier
\newpage
\section{Spectra}
\FloatBarrier
\begin{figure}[ht!]
        \includegraphics[width=\columnwidth]{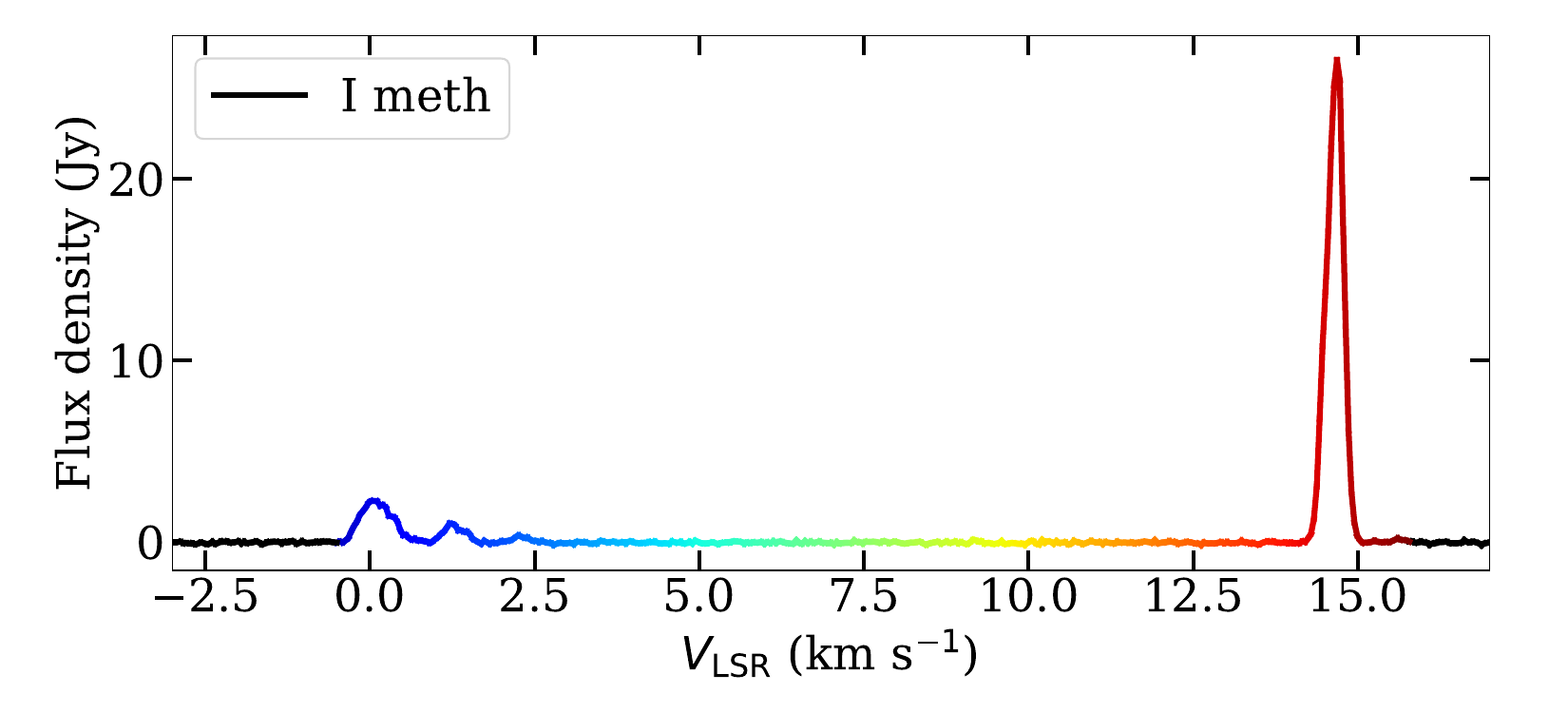}
        \includegraphics[width=\columnwidth]{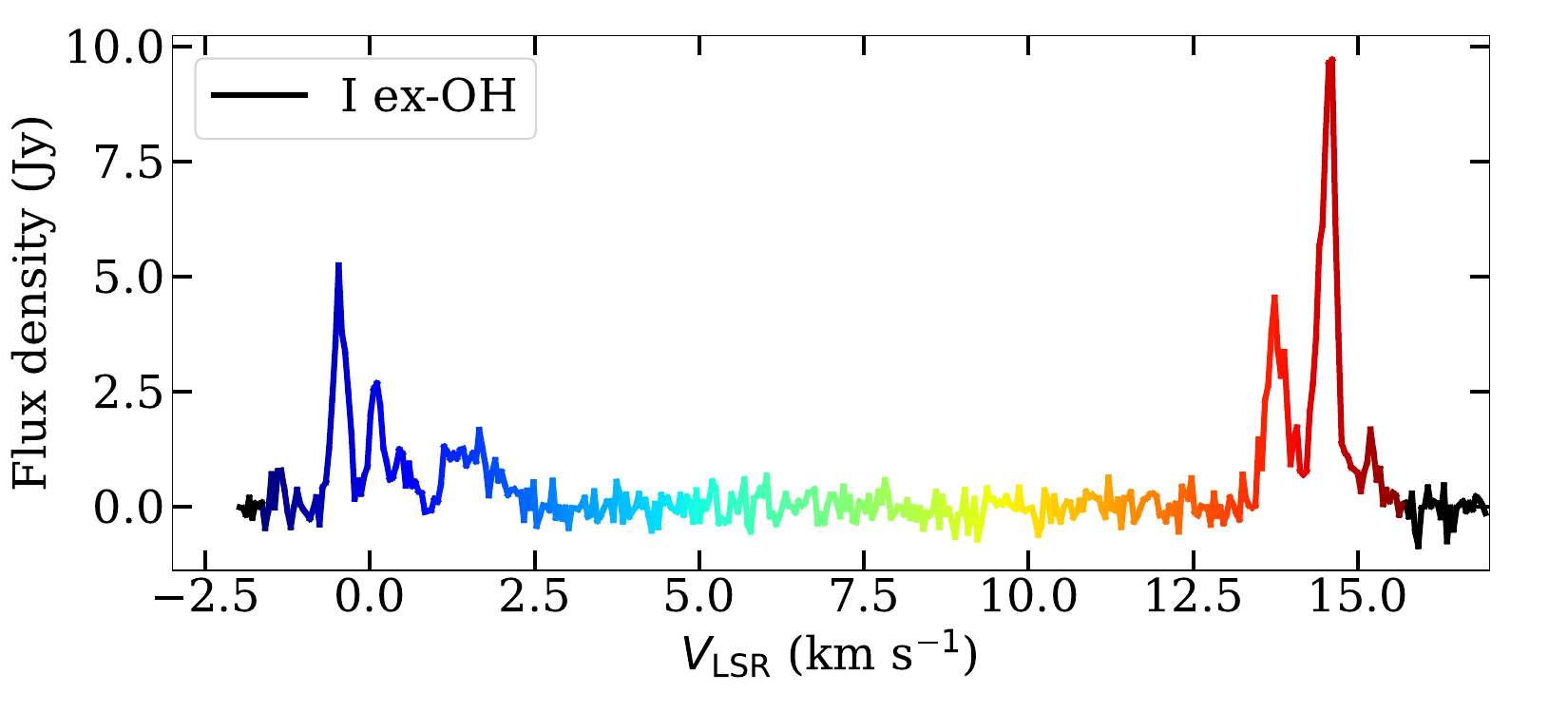}
        \includegraphics[width=\columnwidth]{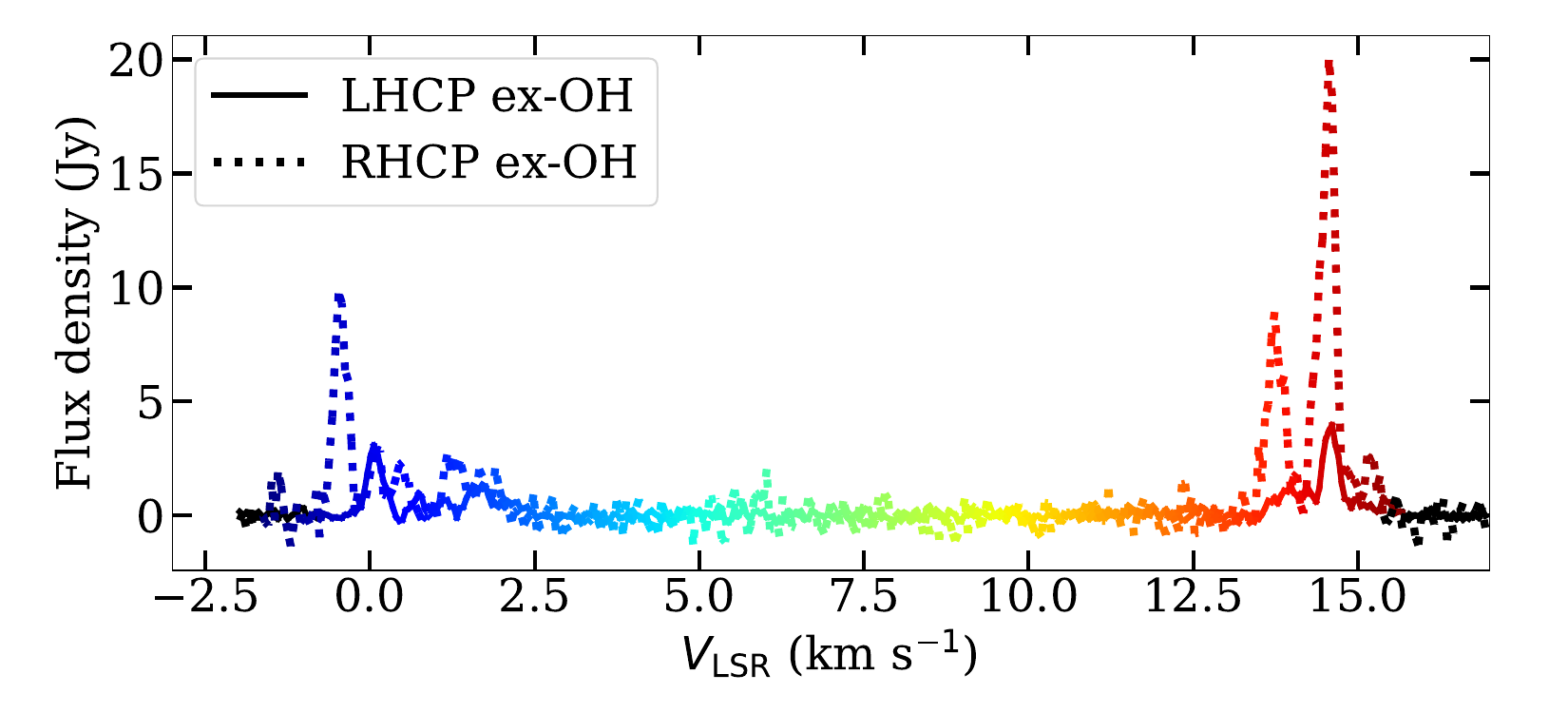}    
    \caption{The total intensity ($I$) spectra of the 6.7\,GHz CH$_3$OH (upper panel) and 6.035 GHz ex-OH (middle panel) maser emissions, and the left- and right-hand circular polarization (LHCP, RHCP) spectra (bottom panel) of the ex-OH maser emission detected toward G69.540-0.976 (ON\,1). The systemic velocity of the region is $V_{\rm{LSR, sys}}^{\rm{ON\,1}}=+11.6\,\rm{km s^{-1}}$ \citep{bronfman1996}.}
    \label{fig:spectra_on1}
    
\end{figure}
\begin{figure}[ht!]
        \includegraphics[width=\columnwidth]{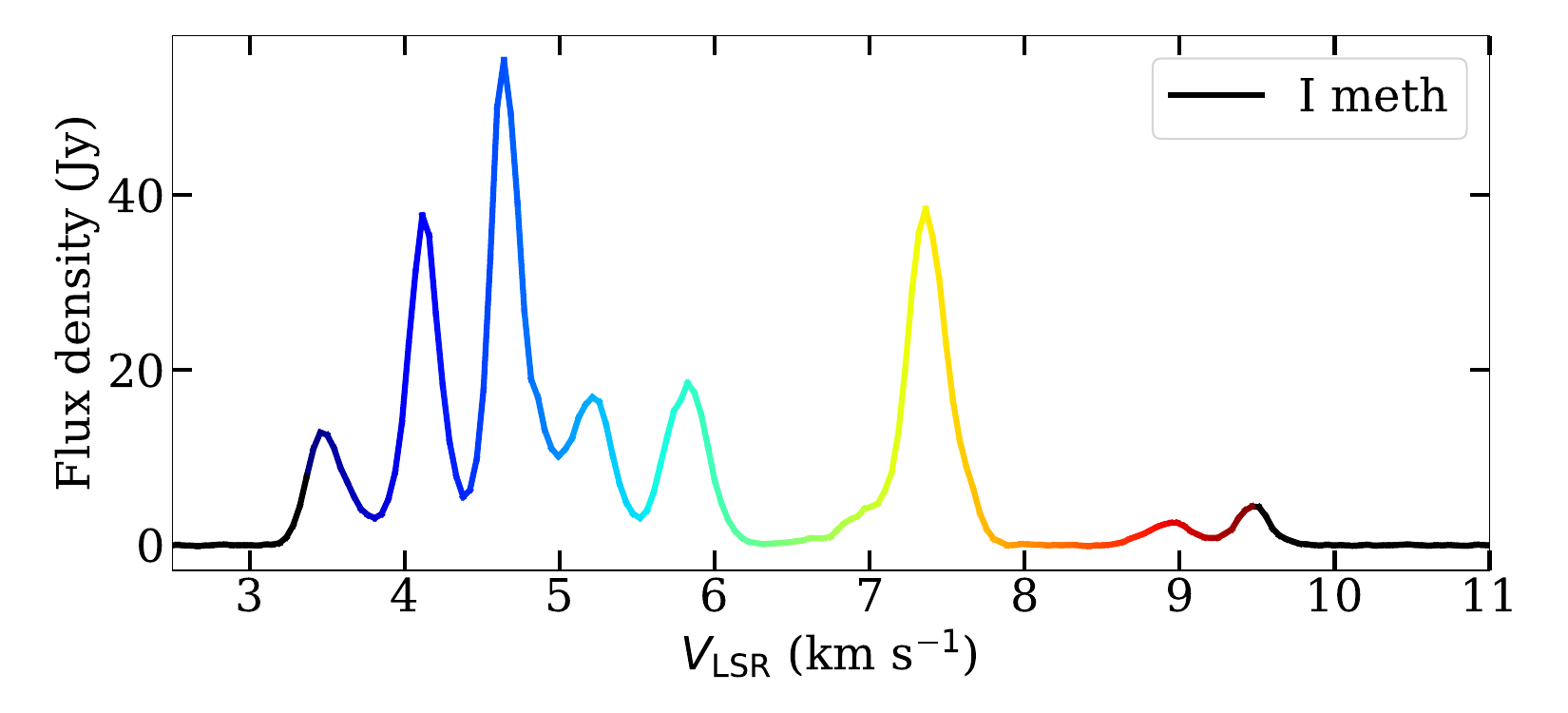}
        \includegraphics[width=\columnwidth]{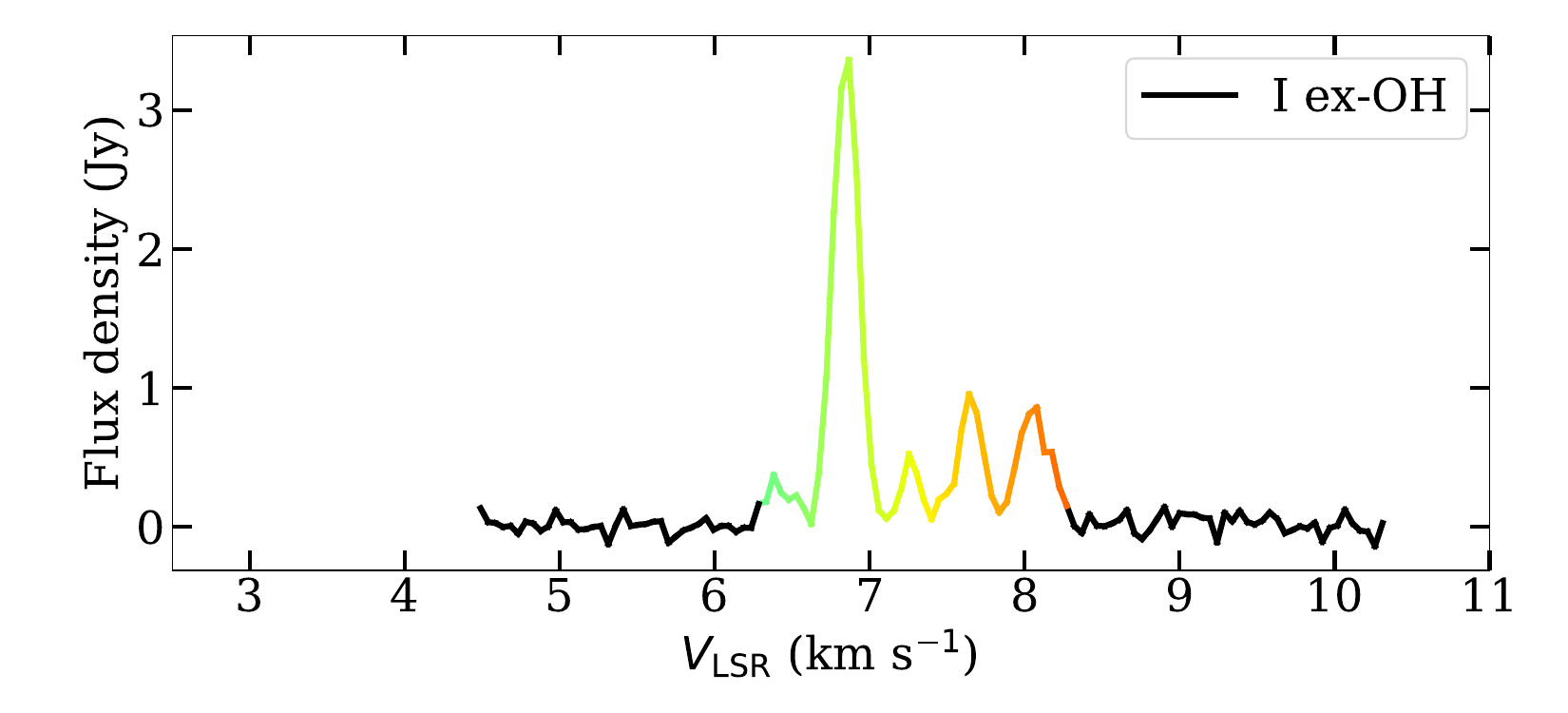}
        \includegraphics[width=\columnwidth]{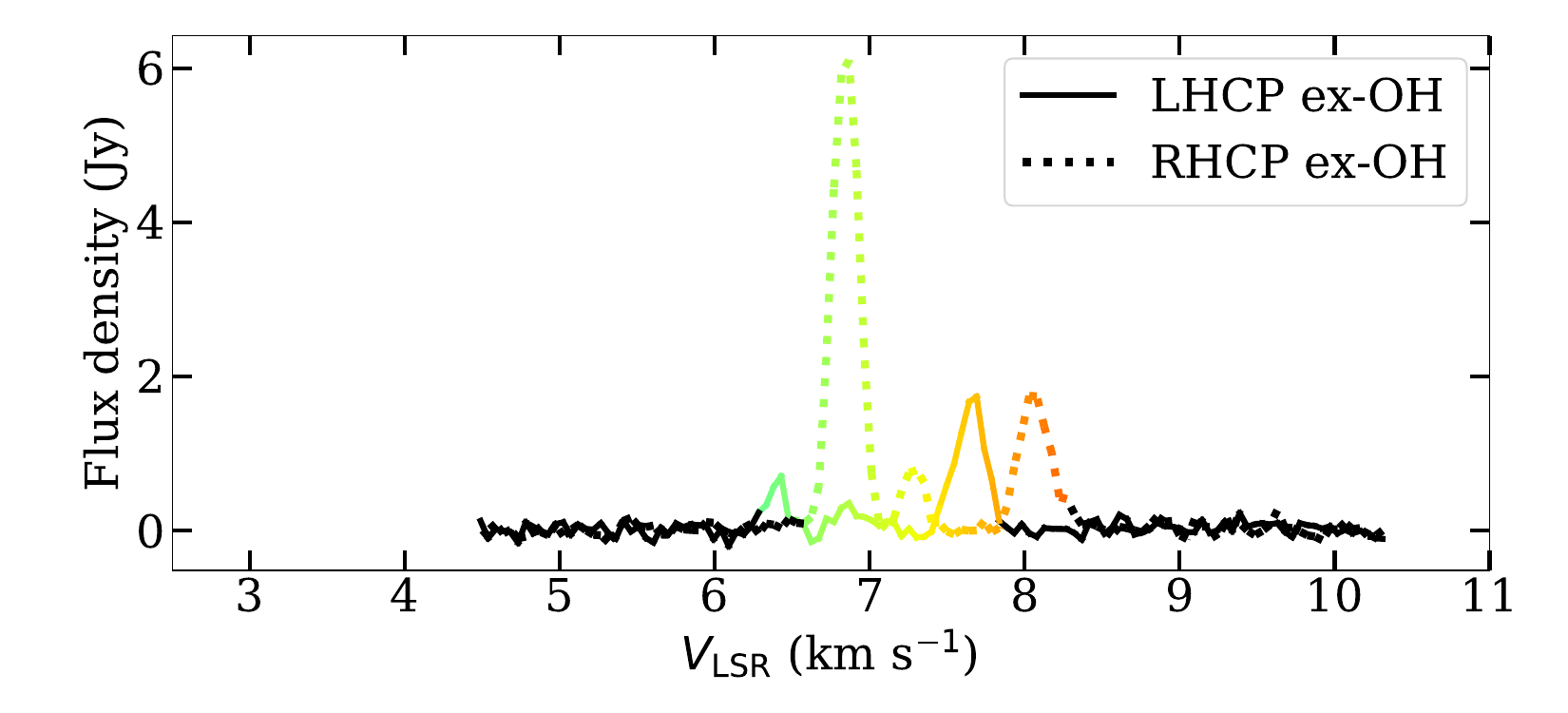}    
    \caption{Similar to Fig.\,\ref{fig:spectra_on1} but for G81.871+0.781 (W75N). The systemic velocity of the region is $V_{\rm{LSR, sys}}^{\rm{W75N}}=+10.0\,\rm{km s^{-1}}$ \citep{shepherd2003}.}
    \label{fig:spectra_w75}
\end{figure}

\end{appendix}

\end{document}